\documentclass[conference]{IEEEtran}
\IEEEoverridecommandlockouts
\usepackage{fancyhdr}
\usepackage{cite}
\usepackage{amsmath,amssymb,amsfonts}
\usepackage{algorithm}
\usepackage{algorithmicx}
\usepackage{algpseudocode}
\usepackage{graphicx}
\usepackage{textcomp}
\usepackage{xcolor}
\usepackage{array}    
\usepackage{fancyhdr}
\usepackage{float}
\usepackage{subcaption}
\usepackage{url}
\usepackage{multirow}
\usepackage{hyperref}

\def\BibTeX{{\rm B\kern-.05em{\sc i\kern-.025em b}\kern-.08em
    T\kern-.1667em\lower.7ex\hbox{E}\kern-.125emX}}

\usepackage{titlesec}

\usepackage{amsmath}

\renewcommand{\thesection}{\arabic{section}}
\renewcommand{\thesubsection}{\arabic{section}.\arabic{subsection}}
\renewcommand{\thesubsubsection}{\arabic{section}.\arabic{subsection}.\arabic{subsubsection}}

\titleformat{\section}[block]{\bfseries\normalsize}{\thesection\quad}{0pt}{}  
\titleformat{\subsection}[block]{\itshape\normalsize}{\thesubsection\quad}{0pt}{}  
\titleformat{\subsubsection}[block]{\itshape\normalsize}{\thesubsubsection\quad}{0pt}{} 

\renewcommand{\baselinestretch}{1}

\begin{document}
\rmfamily

\title{{\fontsize{16}{19}\selectfont \textbf{Optimizing Battery and Line Undergrounding Investments for Transmission Systems under Wildfire Risk Scenarios:\\[-0.55em] A Benders Decomposition Approach}}}


\author{

    \IEEEauthorblockN{{\fontsize{12}{14}\selectfont Ryan Piansky}}
    \IEEEauthorblockA{\textit{\fontsize{10}{12}\selectfont School of Electrical and Comp. Eng.} \\
    \textit{\fontsize{10}{12}\selectfont Georgia Institute of Technology}\\
    \fontsize{10}{12}\selectfont Atlanta, GA, USA \\
    \fontsize{10}{12}\selectfont rpiansky3@gatech.edu}
    \and
   
    \IEEEauthorblockN{{\fontsize{12}{14}\selectfont Rahul Gupta}}
    \IEEEauthorblockA{\textit{\fontsize{10}{12}\selectfont School of Electrical Eng. and Comp. Science} \\
    \textit{\fontsize{10}{12}\selectfont Washington State University}\\
    \fontsize{10}{12}\selectfont Pullman, WA, USA \\
    \fontsize{10}{12}\selectfont rahul.k.gupta@wsu.edu}
    \and
   
    \IEEEauthorblockN{{\fontsize{12}{14}\selectfont Daniel K. Molzahn}}
    \IEEEauthorblockA{\textit{\fontsize{10}{12}\selectfont School of Electrical and Comp. Eng.} \\
    \textit{\fontsize{10}{12}\selectfont Georgia Institute of Technology}\\
    \fontsize{10}{12}\selectfont Atlanta, GA, USA \\
    \fontsize{10}{12}\selectfont molzahn@gatech.edu}
}
\maketitle
\thispagestyle{fancy}

\begin{abstract}
With electric power infrastructure posing an increasing risk of igniting wildfires under continuing climate change, utilities are frequently de-energizing power lines to mitigate wildfire ignition risk, which can cause load shedding. Recent research advocates for installing battery energy storage systems as well as undergrounding risky overhead lines to reduce the load shedding during such de-energizations. Since wildfire ignition risk can exhibit substantial geographic and temporal variations, it is important to plan battery installation and line undergrounding investments while considering multiple possible scenarios. This paper presents a scenario-based framework for optimizing battery installation and line undergrounding investments while considering many scenarios, each consisting of a day-long time series of uncertain parameters for the load demand, renewable generation, and wildfire ignition risks. This problem is difficult to solve due to a large number of scenarios and binary variables associated with the battery placements as well as the lines to be undergrounded. To address the computational challenges, we decompose the problem in a two-stage scheme via a Benders decomposition approach. The first stage is a master problem formulated as a mixed-integer linear programming (MILP) model that makes decisions on the locations and sizes of batteries as well as the lines to be undergrounded. The second stage consists of a linear programming model that assesses these battery and line undergrounding decisions as modeled by a DC optimal power flow formulation. We demonstrate the effectiveness of the proposed scheme on a large transmission network with actual data on wildfire ignition risks, load, and renewable generation. 
\end{abstract}


\vspace{-1em}
\section{Introduction}\label{sec:intro}
With the growing prevalence of severe wildfires, mitigating climate change-driven natural disasters necessitates the development of effective computational methods for planning resilient infrastructure. Worsening climate change coupled with aging equipment is leading to increasingly risky wildfire conditions~\cite{martinuzzi2019future}. Wildfires started by power system infrastructure are not uncommon~\cite{CPUCignitions} and tend to be more severe and expansive when compared to other ignition sources~\cite{keeley2019twenty, syphard2015location}, likely because high wind speeds and temperatures correlate with both increased power line fault probability and fire spread.

Public Safety Power Shutoff (PSPS) events are one method for mitigating the chance that a power line fault will ignite a fire. During PSPS events, a utility preemptively de-energizes certain power lines to remove their ignition risk~\cite{pge2021PSPS}. Utilities identify lines that are at high-risk of starting a fire based on line-specific factors (condition, age, capacity, etc.), environmental conditions (temperature, wind velocity, humidity, etc.), wildfire spread models, and other considerations~\cite{sotolongo2020california}. Lines that exceed an acceptable risk level are de-energized to prevent a fault or sparks from igniting what could be a severe fire.

PSPS events offer an effective and immediate way for utilities to temporarily remove excessively risky lines from operation. However, as lines in a network are de-energized, the utility's overall ability to transmit power is reduced, leading to power outages for consumers. Power outages, including those caused by PSPS events, can have significant economic impacts and negatively affect communities and consumers relying on that power~\cite{wong2022support}. Utilities are in the process of hardening their systems to reduce the extent of PSPS events through line undergrounding, covered conductors, and vegetation management~\cite{pge2023undergrounding,pge2023hardening,sce_wmp,sce_coveredconductor}.

Utilities like PG\&E and SCE in California are currently in the process of undergrounding portions of their transmission network~\cite{Blunt_2023,PGE,sce_wmp}. Line undergrounding can be an effective long-term tool to mitigate wildfire ignition risks while still allowing transmission lines to carry power, thus reducing load shed. However, undergrounding lines is costly (frequently between \$5 and \$10 million per mile) and lengthy timelines are required for these projects~\cite{hall2012out, yang2022resilient, Beam_2023}. 

Utilities are also investing in grid-scale batteries~\cite{yang2018,marnell2019transmission}. These can serve power to local consumers that may be isolated during a PSPS event~\cite{singer2023batteries, murray2022caiso}. While batteries are helpful during these events, they can also benefit grid operation during normal conditions with low wildfire ignition risk (e.g., via price arbitrage or improving renewable integration~\cite{santos2022influence}). 

\subsection{Related work}
Battery sizing problems need to consider multiple wildfire scenarios, as a single scenario may lead to suboptimal decisions since locations subjected to high wildfire ignition risk change depending on the time of year. Previous research has looked at limited multi-scenario battery sizing, siting and operation at the distribution level under periods of high wildfire risk~\cite{bertoletti2023unbalanced}. The authors in \cite{kody2022optimizing} aggregate a ``worst case'' wildfire profile to plan for battery installation with a single scenario. Given the spatio-temporally varying nature of wildfire ignition risk, the differing lines that are de-energized during PSPS events can significantly affect the distribution of the load shedding throughout year and in different scenarios. To address this issue, previous work in \cite{piansky2024long} considers a full year of high and low wildfire ignition risk days when optimizing battery placements on a small test network. The results demonstrated the advantages of considering the full year of data and showed that battery placements differed substantially compared to solutions that only considered a subset of scenarios.

In this paper, we propose a scenario-based stochastic optimization approach where we model a fixed set of days representative of conditions throughout the year. These days consist of a mixture of low-risk and wildfire-prone days. Even when attempting to solve for battery placement and line undergrounding decisions for a fixed set of days, these problems can quickly become intractable. Binary placement decision variables introduce computational challenges that lead to very slow solution times. Such computational difficulties are often addressed by decomposition methods that facilitate parallelization. For example, the authors in \cite{piansky2024long} propose a Progressive Hedging technique to decompose with respect to time. Another widely used approach is Benders decomposition \cite{benderspartitioning, conejo2006decomposition, rahmaniani2017benders}, which facilitates the separation of investment and operational constraints by considering binary investment decisions as complicating variables \cite{conejo2006decomposition}. Benders decomposition has frequently been used for battery planning problems in prior literature. 
For example, references \cite{nick2017optimal} and \cite{ranjbar2021resiliency} use Benders decomposition in the planning of battery storage in distribution systems. In \cite{yi2022expansion} and \cite{yi2022optimal}, Benders decomposition is used for network expansion and line reinforcement in distribution networks. These schemes have been shown to improve the computational speed of MILP formulations that have similar mathematical structure to our planning problem. 

\subsection{Contributions}\label{sec:contributions}
This paper proposes a two-stage algorithm based on Benders decomposition to solve large-scale infrastructure resilience planning problems. The first stage optimizes the battery sizing and siting decisions and the line undergrounding locations. The second stage's subproblem optimizes grid operation for different scenarios after fixing the investment decisions from the master planning problem. This algorithm allows for decomposition with respect to the scenarios.

\begin{figure}[t]
    \centering
	\includegraphics[width=0.95\linewidth]{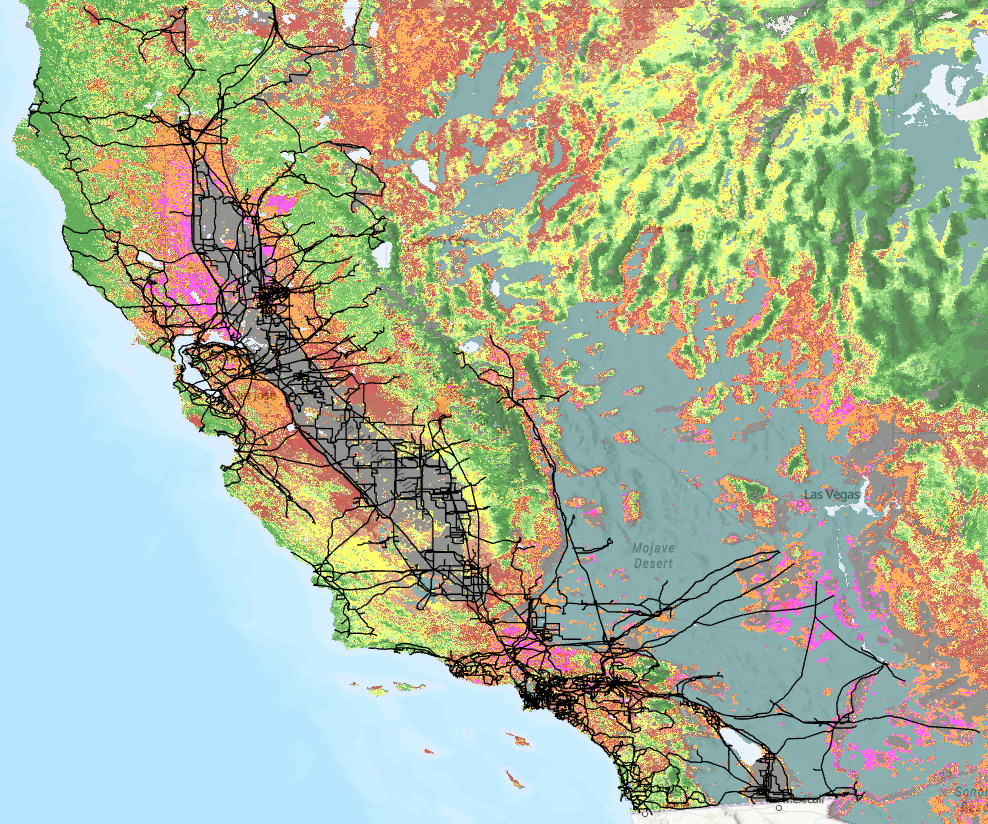}
	\caption{California's transmission line paths on a Wildland Fire Potential Index map from 2020.}
	\label{fig:cats_wfpi_network}       
    \vspace{-1em}
\end{figure}

This algorithm is applied to a large-scale and realistic synthetic transmission network, the California Test System (CATS)~\cite{taylor2023california}. This test system is augmented with real-world hourly renewable and load information~\cite{taylor2023california} as well as daily real-world wildfire ignition risk data~\cite{piansky2025quantifying}. Figure~\ref{fig:cats_wfpi_network} shows the CATS network overlaid on a snapshot of risk values from the United States Geological Survey's (USGS) Wildland Fire Potential Index (WFPI)~\cite{USGS2024Wildland}. To the best of our knowledge, this paper presents the first solution of an infrastructure resilience investment problem at this scale in the academic literature. Our Benders decomposition algorithm allows for the infrastructure decisions to be made in an optimal way for the different scenarios at varying times of the year. We show the benefit of optimally planning for multiple scenarios throughout the year.

The paper is organized as follows. Section~\ref{sec:problem_formulation} formulates the infrastructure resilience planning problem considering the constraints of the transmission system. Section~\ref{sec:decomposition} describes the decomposition approach. 
Section~\ref{sec:case} presents the simulation setup. Section~\ref{sec:results} demonstrates numerical results. Section~\ref{sec:conclusion} concludes the paper.

\section{Problem Formulation}\label{sec:problem_formulation}
This section describes the optimal planning problem for battery sizing, siting, and operation with line undergrounding to mitigate the load shedding associated with wildfire ignition risk mitigation from transmission systems line de-energizations.\footnote{Since prior work in~\cite{kody2022optimizing} found batteries and line undergrounding to be an effective combination when balancing wildfire ignition risk reduction and load shed mitigation from PSPS events, we focus on these infrastructure investments.} The objective function minimizes the sum of the investment cost of battery storage and line undergrounding as well as the cost of load shedding. The planning problem's decisions are the battery sizes and locations as well as the lines that are undergrounded. These decisions are modeled by binary variables, thus making the problem formulation a mixed-integer problem (MIP). The constraints consist of the power flow equations for the transmission system along with operational limits on the voltages and power flows. Using a nonlinear AC power flow model leads to intractable mixed-integer nonlinear programs, so we model the power flow constraints via the B$\theta$ DC power flow linearization, which is often applied when modeling transmission networks. 

In the remainder of this section, we describe the objective function and the investment and operational constraints of the batteries and the transmission network.

\subsection{Objective function}
The objective function consists of five terms: (i) the investment cost of the battery including a fixed installation cost and a cost per MWh of battery capacity, (ii) cost of line undergrounding, (iii) the cost of load shedding, (iv) the cost of generation, and (v) a dummy cost to avoid nonphysical simultaneous battery charging and discharging. 

Let the symbols $X_i^E \in \mathbb{R}^+$,  $X_i^P \in \mathbb{R}^+$, and $X_i^F \in \{0,1\}$ denote the battery decision variables for energy capacity, power rating, and location, respectively. Let the symbols ${C}^E$, ${C}^P$, and ${C}^F$ denote the costs with respect to the energy and power sizes and the fixed investment cost, respectively. Let the symbol $X^{UG}_{ij} \in \{0,1\}$ denote the binary decision variable for the risky lines. Let $C^{UG}$ refer to the fixed cost of undergrounding per mile of a line and $\mathfrak{L}_{ij}$ denote the length of the line. 
We denote the sets of candidate nodes for the battery storage, nodes with non-zero demands, and nodes with generators as $\mathcal{B}$, $\mathcal{D}$, and $\mathcal{G}$, respectively. The set of risky lines is denoted by $\mathcal{L}^{\text{risk}}$.
The sets of scenarios and timesteps in a particular scenario are denoted by $\Omega$ and $\mathcal{T}$, respectively.
For bus $i$, we denote the amount of load shed and generation at time $t$ in scenario $\omega$ by $p^{ls}_{i, t,\omega}$ and $p^{g}_{i, t,\omega}$, respectively. The symbols $c^{ls}_{t,\omega}$ and $c^{g}_{t,\omega}$ denote the costs of load-shedding and generation, respectively, at time $t$ in scenario $\omega$. The symbols $p^{b,ch}_{i, t,\omega}$ and $p^{b,dch}_{i, t,\omega}$ refer to the battery charge and discharge variables for time $t$ in scenario $\omega$.

Then, the objective function is
\begin{align}
\begin{aligned}
   &  \underbrace{\sum_{i \in \mathcal{B}} ({C}^E X_i^E + {C}^P X_i^P + {C}^F X_i^F )}_{\text{Battery investment}}  + \\
   &  \underbrace{\sum_{(i,j) \in \mathcal{L}^\text{risk}} {C}^{UG}  \mathfrak{L}_{ij} X^{UG}_{ij} }_{\text{Line undergrouding investment}}  + \\
   & \sum_{\omega\in\Omega}\sum_{t\in\mathcal{T}}\Bigg(\underbrace{\sum_{i \in \mathcal{D}}c^{ls}_{t,\omega}p^{ls}_{i, t,\omega}}_{\text{Load shedding}} + \underbrace{\sum_{i \in \mathcal{G}}c^{g}_{t,\omega}p^{g}_{i,t,\omega}}_{\text{Generation}}\Bigg)\\
    & \lambda \sum_{\omega\in\Omega}\sum_{t\in\mathcal{T}}\Bigg(\underbrace{\sum_{i \in \mathcal{B}}(p^{b,ch}_{i, t,\omega} + p^{b,dch}_{i,t,\omega})}_{\text{Battery operation cost}}\Bigg)
\end{aligned}
\label{eq:objective}
\end{align}
and is described as follows.
\begin{itemize}
    \item \textbf{Battery investment:} This term consists of the cost per power and energy size of the battery and a fixed installation cost per battery. The total investment cost is the sum of all the batteries placed in the network.
    \item \textbf{Line undergrounding investment:} This term consists of the cost of undergrounding the lines that are in areas of high wildfire ignition risk. The cost is defined as the total cost of undergrounding a subset of the lines within the candidate set of risky lines contained in $\mathcal{L}^\text{risk}$. The set of candidate lines is predetermined based on historical wildfire ignition risk data from \cite{piansky2025quantifying}.
    \item \textbf{Load shedding:} This term refers to the penalty per MWh of the load shed in the network caused by de-energizing lines that are in areas of high wildfire ignition risk. The cost of load shedding can vary greatly depending on the affected consumers' characteristics (e.g., residential vs. industrial). 
    \item \textbf{Generation:} This term refers to the generator operating cost which is modeled by the cost curves from~\cite{taylor2023california}.
    \item  \textbf{Battery operation:} This term refers to a dummy cost that is included in the objective to prevent the battery from being simultaneously charged and discharged for a given time $t$ in the scenario $\omega$. This term is inspired by the previous work in \cite{stai2017dispatching, gupta2022coordinated} which shows that such an objective discourages the charge and discharge variables to be simultaneously nonzero. 
\end{itemize}

\subsection{Constraints}
We next describe the planning problem's constraints related to the operation of the battery and the transmission grid. 
\subsubsection{Battery investment constraints}
Constraints modeling the batteries' energy and power capacities are expressed as
\begin{subequations}
 \label{eq:battery_constraints}
\begin{align}
        & \underline{X}^{P} \leq  X^{P}_i   \leq \overline{X}^{P} && i \in \mathcal{B},  \label{eq:Power_size}\\
        & \underline{X}^{E} \leq  X^{E}_i   \leq \overline{X}^{E} && i \in \mathcal{B},  \label{eq:Energy_size}\\
        & X^{E}_i \leq \overline{X}^{E}_i X^{F}_i, && X^{F}_i\in \{0,1\}, i \in \mathcal{B}. \label{eq:location_cons} 
\end{align}
\end{subequations}
Constraints \eqref{eq:Power_size} and \eqref{eq:Energy_size} limit the power and energy capacities and constraint \eqref{eq:location_cons} ensures that batteries are only sized at nodes chosen as battery installation locations.

\subsubsection{Battery operational constraints}
Now, we describe the operational constraints of the battery, namely, the operating limits on the power and energy. 
Denote the State-of-Energy for the battery at bus $i$ as SoE$_{i,t,\omega}$ for time~$t$ and scenario $\omega$. The SoE$_{i,t,\omega}$ is related to the charging and discharging variables as follows:

\begin{align}
\begin{aligned}
	\text{SoE}_{i,t+1,\omega} = \gamma(\text{SoE}_{i,t,\omega}) + \eta \frac{p^{b, ch}_{i, t,\omega} \Delta t}{3600}  - \frac{1}{\eta} \frac{p^{b,dch}_{i, t,\omega} \Delta t}{3600}, \\ \quad i\in\mathcal{B}, t\in\mathcal{T}, \omega \in \Omega,
\end{aligned} \label{eq:soc}
\end{align}
where $\Delta t$ refers to the period between two subsequent timesteps and the symbol $\gamma$ is a parameter (close to 1) that models the hourly self-discharge of the battery. The expressions $\eta$ and $\frac{1}{\eta}$ are the efficiency of battery charging and discharging.

The constraints on the SoE$_{i,t,\omega}$ are
\begin{align}
	& \alpha X^{E}_i \leq \text{SoE}_{i,t,\omega} \leq (1-\alpha)X^{E}_i \quad i\in\mathcal{B}, t\in\mathcal{T}, \omega \in \Omega,
\end{align}
where $\alpha = 0.1$ is a safety factor that is generally used to avoid deep discharge or over charge of the battery storage.

We also have limits on the battery power defined as
\begin{subequations}
\begin{align}
	& 0 \leq p^{b, ch}_{i,t,\omega} \leq X^{P}_i, \quad i\in\mathcal{B}, t\in\mathcal{T}, \omega \in \Omega, \\
    & 0 \leq p^{b,dch}_{i,t,\omega} \leq X^{P}_i, \quad i\in\mathcal{B}, t\in\mathcal{T}, \omega \in \Omega.
\end{align}
\end{subequations}
Finally, we enforce the following constraint
\begin{align}
        0 \leq p^{b, ch}_{i,t,\omega} + p^{b,dch}_{i,t,\omega} \leq X^{P}_i \quad i\in\mathcal{B}, t\in\mathcal{T}, \omega \in \Omega
\end{align}
that also helps mitigate simultaneous charging and discharging of batteries as discussed in \cite{marley2016solving, pozo2022linear}. 
\subsubsection{Transmission grid operational constraints} 
We model the transmission network constraints using the DC power flow approximation. To model the risk of wildfire ignition from the transmission lines, we follow the approach deployed in~\cite{piansky2025quantifying} where each transmission line in the network is assigned a daily risk of wildfire ignition. Likewise, we consider that the lines above a certain wildfire ignition risk are switched off to remove the risk of wildfire ignition from those lines. Let $\mathcal{L}^\text{on}_\omega$ denote the lines that are safe to operate according to the predetermined risk threshold for a scenario $\omega$ of the daily wildfire ignition risk profile.  
Let $f^{ij}_{t,\omega}$ denote the power flow for (time, scenario) $=(t,\omega)$ on the line between nodes $i$ and $j$ and define flow limits of $[-\overline{f}^{ij}~\overline{f}^{ij} ]$. Let $b^{ij}$ denote the line susceptance for the line between nodes $i$ and $j$.
Then, the DC power flow constraints are
\begin{subequations}
\begin{align} 
\nonumber & \forall (i,j) \in \mathcal{L}_{\omega}^{\text{on}}, \forall t \in \mathcal{T}, \forall \omega \in \Omega :\\
        & \qquad -\overline{f}^{ij} \leqslant f^{ij}_{t,\omega} \leqslant \overline{f}^{ij},\\
        & \qquad f_{t,\omega}^{ij} = -b^{ij}(\theta_{t,\omega}^i - \theta_{t,\omega}^j).
\end{align}
Let $\theta_{t,\omega}^{n}$ denote the voltage angle for a bus $n$. The angle difference across the line from node $i$ to node $j$ is bounded by the limits $[\underline{\delta}^{ij},~\overline{\delta}^{ij}]$:
\begin{align}
            \underline{\delta}^{ij}  \leqslant \theta^{i}_{t,\omega} - \theta^{j}_{t,\omega}  \leqslant \overline{\delta}^{ij}, && \forall (i,j) \in \mathcal{L}_{\omega}^{\text{on}}, \forall t \in \mathcal{T}, \forall \omega \in \Omega.
\end{align}
\end{subequations}

We then have constraints for the lines that are candidates for undergrounding. We model the undergrounding decision by the binary variables $X_{ij}^{UG} \in \mathcal{L}^\text{risk}$, where $\mathcal{L}^\text{risk}$ defines the set of lines that are considered as candidates for undergrounding. This set contains the union of the risky lines for each scenario: $\mathcal{L}^\text{risk} = \bigcap_{\omega \in \Omega} \mathcal{L}_{\omega}\setminus\mathcal{L}_{\omega}^{\text{on}}$. The constraints for undergrounded lines are formulated using a big-M method since the lines to be undergrounded are decision variables in the problem. The big-M constant is denoted by $M$ and is tuned according to approach described in \cite{piansky2025quantifying}. The constraints are
\begin{subequations}
\begin{align} 
\nonumber & \forall (i,j) \in \mathcal{L}^{\text{risk}}\setminus\mathcal{L}_{\omega}^{\text{on}}, \forall t \in \mathcal{T}, \forall \omega \in \Omega:\\
    & \quad f^{ij}_{t,\omega} \geqslant -\overline{f}^{ij} X_{ij}^{UG}, \\
    &  \quad f^{ij}_{t,\omega}  \leqslant \overline{f}^{ij} X_{ij}^{UG}, \\ 
    & \quad \theta^i_{t,\omega} - \theta^j_{t,\omega}  \leqslant \overline{\delta}^{ij} + M(1-X_{ij}^{UG}),  \\ 
    & \quad \theta^i_{t,\omega} - \theta^j_{t,\omega}  \geqslant \underline{\delta}^{ij} - M(1-X_{ij}^{UG}),  \\ 
    & \quad f_t^{ij} \leqslant -b^{ij}(\theta_{t,\omega}^i - \theta_{t,\omega}^j) +  |b^{ij}|M(1- X_{ij}^{UG}), \\
    & \quad f_t^{ij} \geqslant -b^{ij}(\theta_{t,\omega}^i - \theta_{t,\omega}^j) -  |b^{ij}|M(1- X_{ij}^{UG}),
\end{align}
\end{subequations}
Generator limits are expressed as 
\begin{align}
    \underline{p}^{g}_i \leqslant p^{g}_{i,t,\omega} \leqslant \overline{p}^{g}_i, && \forall i \in \mathcal{G}, \forall t \in \mathcal{T}, \forall \omega \in \Omega, 
    \label{const: gen limits}
\end{align}
and load shedding is bounded by the demand present at each node $p^{d}_{n,t,\omega}$, i.e.,
\begin{align}
    0 \leqslant p^{ls}_{t,\omega} \leqslant p^{d}_{n,t,\omega} , && \forall n \in \mathcal{N}, \forall t \in \mathcal{T},\forall \omega \in \Omega. 
\end{align}
Finally, the nodal active power balance constraints are
\begin{align}
     \begin{aligned}
        \sum_{(i,j) \in \mathcal{L}^{i,\text{fr}}} f^{ij}_{t,\omega} &  - \sum_{(i,j) \in \mathcal{L}^{i, \text{to}}} f^{ij}_{t,\omega}  = \sum_{i \in \mathcal{G}^i} p^{g}_{i,t,\omega} - \\ \sum_{i \in \mathcal{D}^i}p^{d}_{i,t,\omega} + 
        & \sum_{i \in \mathcal{D}^i}p^{ls}_{i,t,\omega} + 
        \sum_{i \in \mathcal{B}^i}(p^{{b,ch}}_{i,t,\omega} - p^{{b,dch}}_{i,t,\omega}), \\ 
        & \forall i \in \mathcal{N}, \ \forall t \in \mathcal{T}, \omega \in \Omega,
    \end{aligned} \label{eq:power_balance}
\end{align}
where $\mathcal{L}^{i,\text{fr}}$ and $\mathcal{L}^{i,\text{to}}$ refer to the sets of lines that originate and terminate at node $i$, respectively. The symbols $\mathcal{G}^i$, $\mathcal{D}^i$, and $\mathcal{B}^i$ are the sets of generator, demand, and battery indices, respectively, at node $i$.

\subsection{Optimal Planning Problem}
Having defined the objective and constraints, the final planning problem is
\begin{align}
\begin{aligned}
    & \text{minimize} \quad \quad \eqref{eq:objective} \\
    & \text{subject to:} \quad \eqref{eq:battery_constraints}\text{--}\eqref{eq:power_balance}.
\end{aligned}
\label{eq:OP}
\end{align}

The optimization problem in \eqref{eq:OP} is a MILP that is difficult to solve and can be intractable due to the large number of binary variables associated with the battery siting and line undergrounding decisions. The time coupling constraint of the battery model also introduces complexity in the problem formulation. In addition, we also would like to solve the problem with several scenarios of wildfire ignition risk defined by the scenario set $\Omega$, which introduces further computational challenges. 

Given the above-mentioned difficulties in the original formulation of the problem \eqref{eq:OP}, we propose applying the Benders decomposition approach described in the next section.



\section{Reformulation using Benders Decomposition}\label{sec:decomposition}
We formulate a scenario-based stochastic optimization problem using a Benders decomposition method where the master problem optimally sizes and sites the battery storage and the undergrounded lines and the subproblems evaluate the optimality of decisions for the energy storage sizes and locations and the planned underground lines through operational modeling. We next describe the decomposition of the planning problem in \eqref{eq:OP} using a Benders decomposition approach \cite{benderspartitioning}.

We define the vectors of variables 
\begin{itemize}
    \setlength\itemsep{0.5em}
    \item $\mathbf{X} = [X_i^P, X_i^E, X_i^F, \forall i\in\mathcal{B}, X_{ij}^{UG}, \forall(i,j)\in\mathcal{L}^\text{risk}]$
    \item $\mathbf{x}_{t,\omega} = [p^{ls}_{i,t,\omega}, p^{g}_{i,t,\omega}, p^{b,ch}_{i,t,\omega}, p^{b,dch}_{i,t,\omega}, \forall i, t, \omega]$
\end{itemize} 
which contain the investment and operational decisions, respectively. 

Using these vectors, the planning optimization problem in \eqref{eq:OP} can be re-written as follows: 
\begin{subequations}
\begin{align}
    \underset{\mathbf{X}, \mathbf{x}_{t,\omega}}{\text{minimize}}~ \underbrace{\mathbf{C}^\top \mathbf{X}}_{\text{Investment cost}} + \underbrace{\sum_{\omega \in \Omega}\sum_{t\in \mathcal{T}} \mathbf{c}_{t,\omega}^\top \mathbf{x}_{t,\omega}}_{\text{Operation cost}}  
\end{align}
subject to:
\begin{align}
	& \mathbf{A}_{t,\omega}\mathbf{X} + \mathbf{a}_{t,\omega}\mathbf{x}_{t,\omega} \leq \mathbf{d}_{t,\omega}, && \forall t\in\mathcal{T}, \omega\in\Omega,\\
	& \mathbf{E}_{t,\omega}\mathbf{X} + \mathbf{e}_{t,\omega}\mathbf{x}_{t,\omega} =  \mathbf{f}_{t,\omega}, && \forall t\in\mathcal{T}, \omega\in\Omega, \\
    &     \mathbf{G}\mathbf{X} \leq \mathbf{g},
\end{align}
\end{subequations}
where $\mathbf{C}$ and $\mathbf{c}_{t,\omega}$ are the vectors for investment cost and operational cost. The symbols $\mathbf{A}_{t,\omega}, \mathbf{E}_{t,\omega}$ and $\mathbf{a}_{t,\omega}, \mathbf{d}_{t,\omega}, \mathbf{e}_{t,\omega}, \mathbf{f}_{t,\omega}$ are appropriate matrices and vectors for the operational constraints \eqref{eq:soc}--\eqref{eq:power_balance}. Likewise, $\mathbf{G}$ and $\mathbf{g}$ are the appropriate matrix and vector for the investment constraints \eqref{eq:battery_constraints}.

\begin{figure}[!htbp]
    \centering
    \includegraphics[width=0.95 \linewidth]{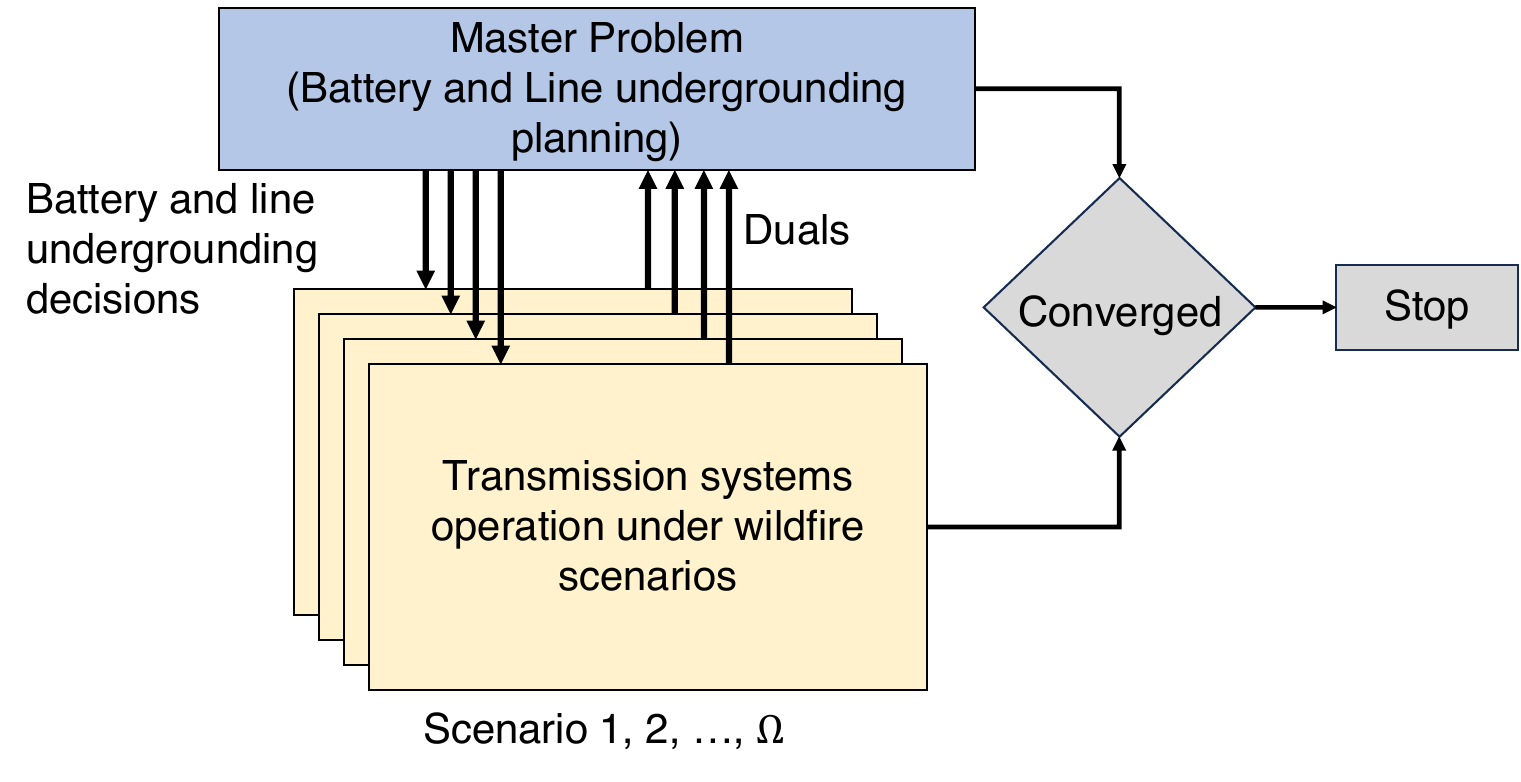}
    \caption{Schematic flow diagram for the Benders decomposition based planning problem.}
    \label{fig:benders_flow}
\end{figure}

We decompose the planning problem by separating the investment variables $\mathbf{X}$ and the operational variables $\mathbf{x}_{t,\omega}$ using the Benders decomposition framework, which divides the optimization problem into a single master problem that minimizes the net investment cost as well as several subproblems that minimize operational costs in various uncertain scenarios. This decomposition facilitates the scalability of the optimization problem as the sub-problems are decomposed with respect to the scenario $\omega$ and can be solved in parallel. The Benders decomposition framework is shown schematically in Fig.~\ref{fig:benders_flow}. The master problem passes the investment decisions to the subproblem. The subproblems then evaluate those decisions (in parallel across scenarios) and pass back the duals corresponding to the constraints imposed for the investment decisions in the subproblems. This process is repeated until convergence. We next detail the subproblems and the master problem.

\subsection{Subproblems} For each scenario $\omega\in\Omega$, we solve an operational problem, referred to as a subproblem, which evaluates the investment decisions provided by the master problem (defined in the next subsection). The subproblem for scenario $\omega$ and for iteration $k$ is formulated as
\begin{subequations}
\label{eq:SP}
\begin{align}
 	\underset{\mathbb{X},\mathbf{x}_{t,\omega}}{\text{minimize}}~ \sum_{t\in \mathcal{T}} \mathbf{c}_{t,\omega} ^\top \mathbf{x}_{t,\omega}  
\end{align}
subject to:
\begin{align}
	& \mathbf{A}_{t,\omega}{\mathbb{X}} + \mathbf{a}_{t,\omega}\mathbf{x}_{t,\omega} \leq \mathbf{d}_{t,\omega},  && \forall t\in\mathcal{T},\\
	& \mathbf{E}_{t,\omega}{\mathbb{X}} + \mathbf{e}_{t,\omega}\mathbf{x}_{t,\omega} =  \mathbf{f}_{t,\omega},  && \forall t\in\mathcal{T},\\
    & \mathbb{X} = \mathbf{\hat{X}}(k-1)  : (\boldsymbol{\nu}_{\omega}(k)). \label{eq:impose_master_decisions}
\end{align}
\end{subequations}
Here, $\mathbb{X}$ represents the continuous relaxed form of the investment decision variables $\mathbf{X}$. This relaxation is implemented to make the subproblem a linear program (LP) by modeling the binary decision variables within $\mathbf{X}$ (such as $X_i^F$ and $X_{ij}^{UG}$) as continuous in the subproblem. The condition in \eqref{eq:impose_master_decisions} ensures that these binary variables ultimately take values of 0 or 1, as dictated by the master solutions. This decomposition into a master problem and subproblems provides the benefit of this relaxation. Additionally, this decomposition allows the subproblems to be solved in parallel for each scenario $\omega \in \Omega$.

As illustrated, we included an extra constraint $\mathbf{X} = \mathbf{\hat{X}}(k-1)$ to enforce the decisions from the master problem at the $(k-1)\textsuperscript{th}$ iteration. Here, $\mathbf{\hat{X}}(k-1)$ pertains to optimizing the battery storage size and location, along with determining which lines should be placed underground. Notably, due to the DC power flow yielding linear grid constraints and the absence of binary variables, the sub-problem in \eqref{eq:SP} is linear.

\subsection{Master problem}
The master problem minimizes the net investment cost of installing batteries and line undergrounding represented by $ \mathbf{C}^\top \mathbf{X}$ and an auxiliary cost $\sum_{\omega \in \Omega} \mathcal{Z}_{\omega}$ representing the subproblem cost. Here, $\mathcal{Z}_{\omega}$ are auxiliary variables which refer to the lower bound of the solution of the subproblem for scenario~$\omega$. Following Benders decomposition, the variable $\mathcal{Z}_{\omega}$ is used to express the Benders cut utilizing the duals of the linking constraint in the subproblem. The master problem is
\begin{subequations}
\label{eq:MP}
\begin{align}
    \mathbf{\hat{X}}(k+1) = \underset{\mathbf{X}}{\text{arg\,min}}~ \mathbf{C}^\top \mathbf{X} + \sum_{\omega \in \Omega} \mathcal{Z}_{\omega}
\end{align}
subject to: 
\begin{align}
     &  \mathbf{G}\mathbf{X} \leq \mathbf{g}, \label{eq:investement_const}\\
     & \mathcal{Z}_{\omega} \geq  \boldsymbol{\hat{\nu}}_{\omega}(k)^\top \big(\mathbf{X} - \mathbf{\hat{X}}(k)\big), && \forall \omega\in\Omega. \label{eq:benders-cut}
\end{align}
Here,~\eqref{eq:investement_const} expresses the investment constraint linked to the battery storage and \eqref{eq:benders-cut} expresses the Benders optimality cut for each scenario $\omega$. The symbol $\boldsymbol{\hat{\nu}}_\omega(k)$ refers to the duals obtained from the $k$--th iteration of the subproblem and for the subproblem associated with scenario $\omega\in\Omega$.\footnote{Symbols with $\hat{\cdot}$ refer to the optimized values.}
\end{subequations}

\subsection{Convergence} The master problem and subproblems are solved iteratively until convergence is attained. The convergence criterion is that the relative difference between the upper and lower bounds on the cost is below a tolerance limit, $\epsilon$. Formally, the upper bound (UB) and the lower bound (LB) on the cost at iteration~$k$ are:
\begin{subequations}
\begin{align}
& \text{LB}_{k} = \mathbf{C}^\top \hat{\mathbf{X}}(k-1)  + \sum_{\omega \in \Omega}\hat{\mathcal{Z}}_{\omega}(k-1),\\
& \text{UB}_{k} = \mathbf{C}^\top \hat{\mathbf{X}}(k-1) + \sum_{\omega \in \Omega}\sum_{t\in \mathcal{T}} \mathbf{c}_{t,\omega} ^\top \hat{\mathbf{x}}_{t,\omega}(k).
\end{align}
\end{subequations}

The Benders decomposition algorithm is summarized in Algorithm~\ref{alg:algo}.
\begin{algorithm}\label{algo}
\caption{Benders Decomposition Algorithm}\label{alg:algo}
\begin{algorithmic}[1]
\State Iteration index, $k$ = 1
\State Solve the master problem \eqref{eq:MP} without Benders cut, i.e., excluding~\eqref{eq:benders-cut}
\State \textbf{repeat}
\If {$(k > 1)$}
\State Solve the master problem \eqref{eq:MP} including a Benders cut constructed using the duals ($\boldsymbol{\nu}_{\omega}$) from the previous iteration of the subproblem.
\EndIf
\For{each scenario $\omega\in\Omega$}
\State Using the optimized $\hat{\mathbf{X}}(k)$ from the master problem, solve the subproblem \eqref{eq:SP} for each scenario  $\omega\in\Omega$ and compute the duals $\boldsymbol{\nu}_{\omega}(k)$.
\EndFor
\State $k\gets k+1$
\State \textbf{until} $\frac{\text{UB}_k-\text{LB}_k}{\text{UB}_k}\leq \epsilon$
\State \textbf{return}
\end{algorithmic}
\end{algorithm}




\section{Case Study}\label{sec:case}

\begin{figure*}[!ht]
\centering
\begin{subfigure}{.48\textwidth}
  \centering
  \captionsetup{width=.9\linewidth}
  \includegraphics[width=.99\linewidth]{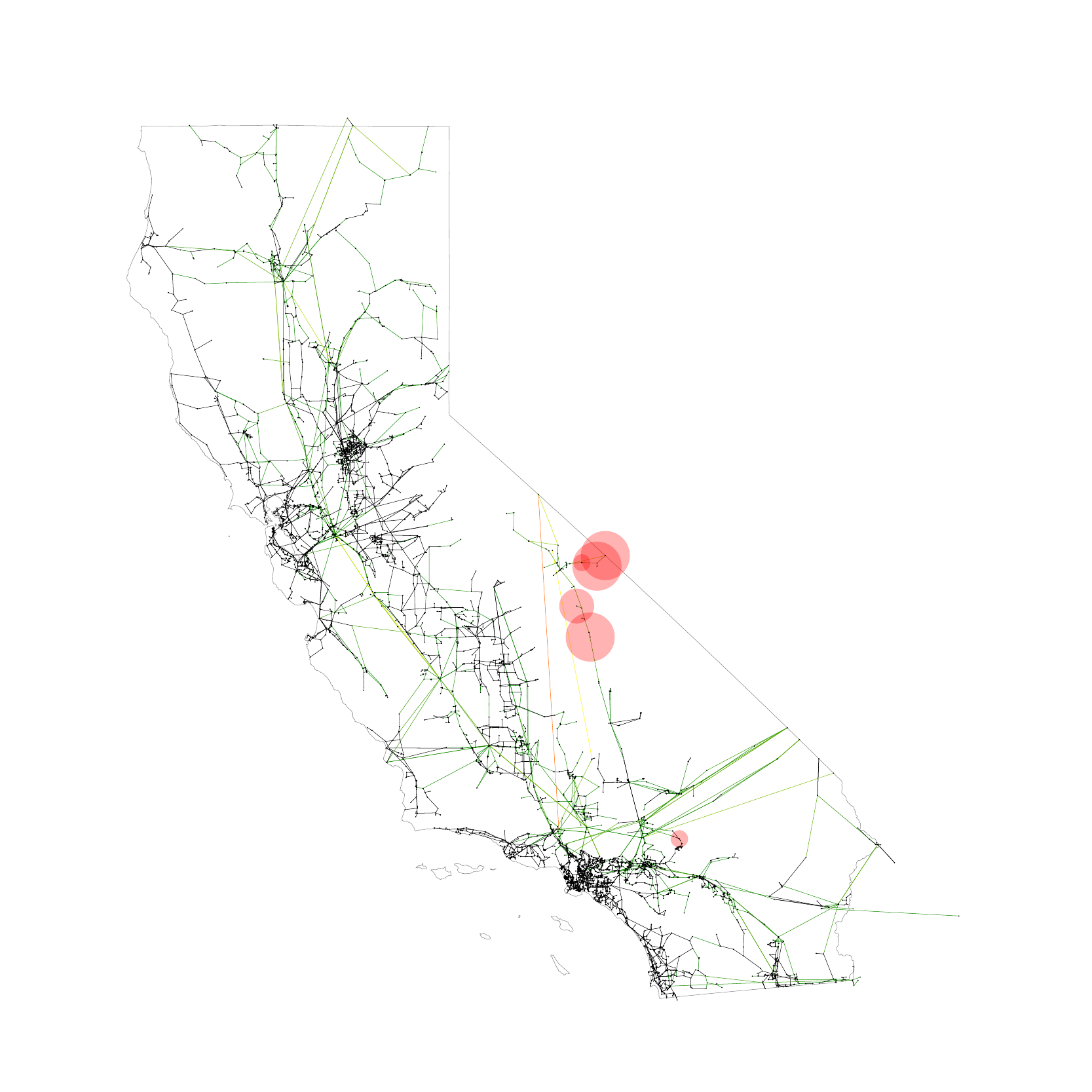}  
  \caption{Battery-only placement results (scheme 1) for the spring season. The size of the circle corresponds to the capacity of the installed battery with the largest circle equating to 400~MWh of installed capacity.}
  \label{subfig:spring_batts}
\end{subfigure}
\begin{subfigure}{.48\textwidth}
  \centering
  \captionsetup{width=.9\linewidth}
  \includegraphics[width=.99\linewidth]{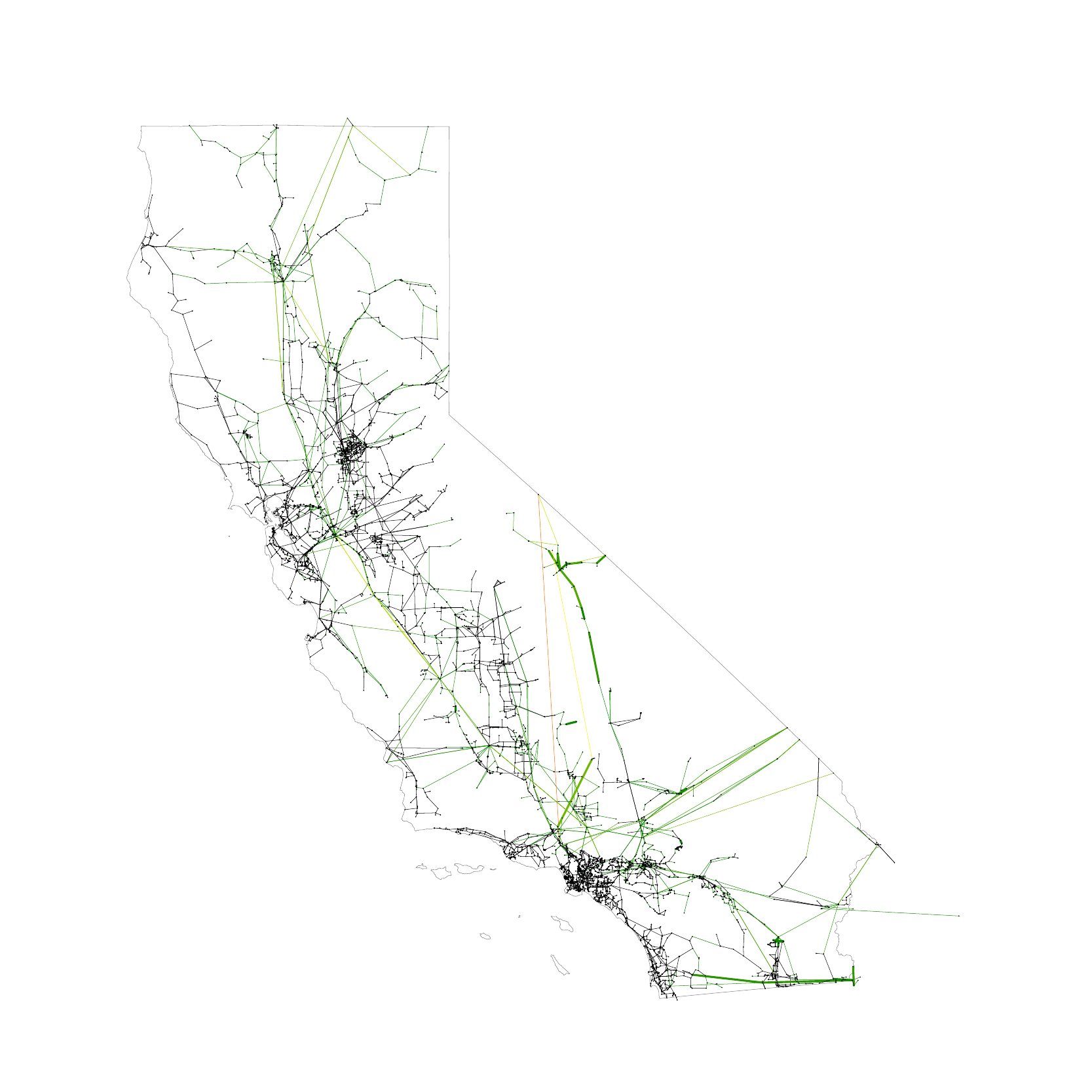}  
  \caption{Battery and undergrounding placement results (scheme 2) for the spring season. Here, no batteries are installed. Thicker lines correspond to transmission lines that have been selected for undergrounding.}
  \label{subfig:spring_batts_ug}
\end{subfigure}
\caption{Optimal battery placements on the CATS network in Spring. Red circles are sized proportionally to the number of batteries placed at that bus with the largest circle representing 400~MWh of installed capacity. Thicker lines represent undergrounded transmission lines. Black lines represent transmission lines that always have zero risk values during the simulated scenarios within the season. All remaining lines have a color representing their maximum risk across the scenarios, with green lines representing lower risk and red lines representing higher risk.}
\label{figure:spring_results}
\end{figure*}

For the numerical results in this paper, we evaluate the proposed planning framework on the CATS network~\cite{taylor2023california}. As described in Section~\ref{sec:contributions}, CATS is a large-scale and realistic synthetic representation of the power transmission network for the state of California. This synthetic transmission network consists of 8870 buses and 10823 lines. The load demanded and renewable generation available at each bus are updated in each scenario based on daily real-world data~\cite{taylor2023california}. 

Transmission lines are assigned a unitless risk value and de-energized based on the threshold method discussed in~\cite{piansky2025quantifying}. Specifically, we use the high-risk cumulative method from~\cite{piansky2025quantifying} to assign risks to transmission lines. De-energization decisions are made based on the 95\textsuperscript{th} percentile from the same paper. This provides a set of lines that will be de-energized based on the historical range of risk associated with each line. Note that this means the line energization statuses, $\mathcal{L}^{\text{risk}}$, is a  parameter and $\mathcal{L}_{\omega}^{\text{on}}$ is a parameter in each scenario.\footnote{Past work has looked at line undergrounding planning problem in conjunction with daily optimal de-energization decisions but this leads to greatly increased computation time\cite{pollack2024equitably}.} All wildfire ignition risk data used for the results in this paper are from 2020.

We set the cost of load shed at $\mathbf{c}_{t,\omega} = $ \$20,000 per MWh based on lower values for the cost of load shed in industrial markets and upper values for cost of load shed in private consumer markets from the values in~\cite{schroder2015value}. We assume that the cost of load shed does not vary during the day or across scenarios. The cost for undergrounding a transmission line is set at $C^{UG} = $ \$7,000,000 per mile~\cite{cpuc_undergrounding}. The fixed cost for a battery installation is  $C^F =$ \$100,000 per node where a battery is placed with a cost of $C^E = $  \$1,000,000 per MWh and $C^{P} = $  \$1,000,000 per MW of installed capacity~\cite{viswanathan20222022}. The fixed costs are divided by the lifetime of the investment to provide an annualized cost. These are then further divided to provide a daily average cost.

To account for the battery losses, we consider efficiency $\eta = 0.95$ and an hourly self-discharge coefficient $\gamma = 0.999958$. These values are derived from \cite{luo2015overview}. For the simulations, we impose the power and energy sizes to be equal ($X^{E}_i  = X^{P}_i$) which means that the battery can be charged from empty to full in an hour. For the limits on the power/energy sizes per node, we consider 4.0 p.u. corresponding to a maximum size battery of 400MWh/400MW at a given node.

\section{Results}\label{sec:results}
\begin{figure*}[!htbp]
\centering
\begin{subfigure}{.48\textwidth}
  \centering
  \captionsetup{width=.9\linewidth}
  \includegraphics[width=.99\linewidth]{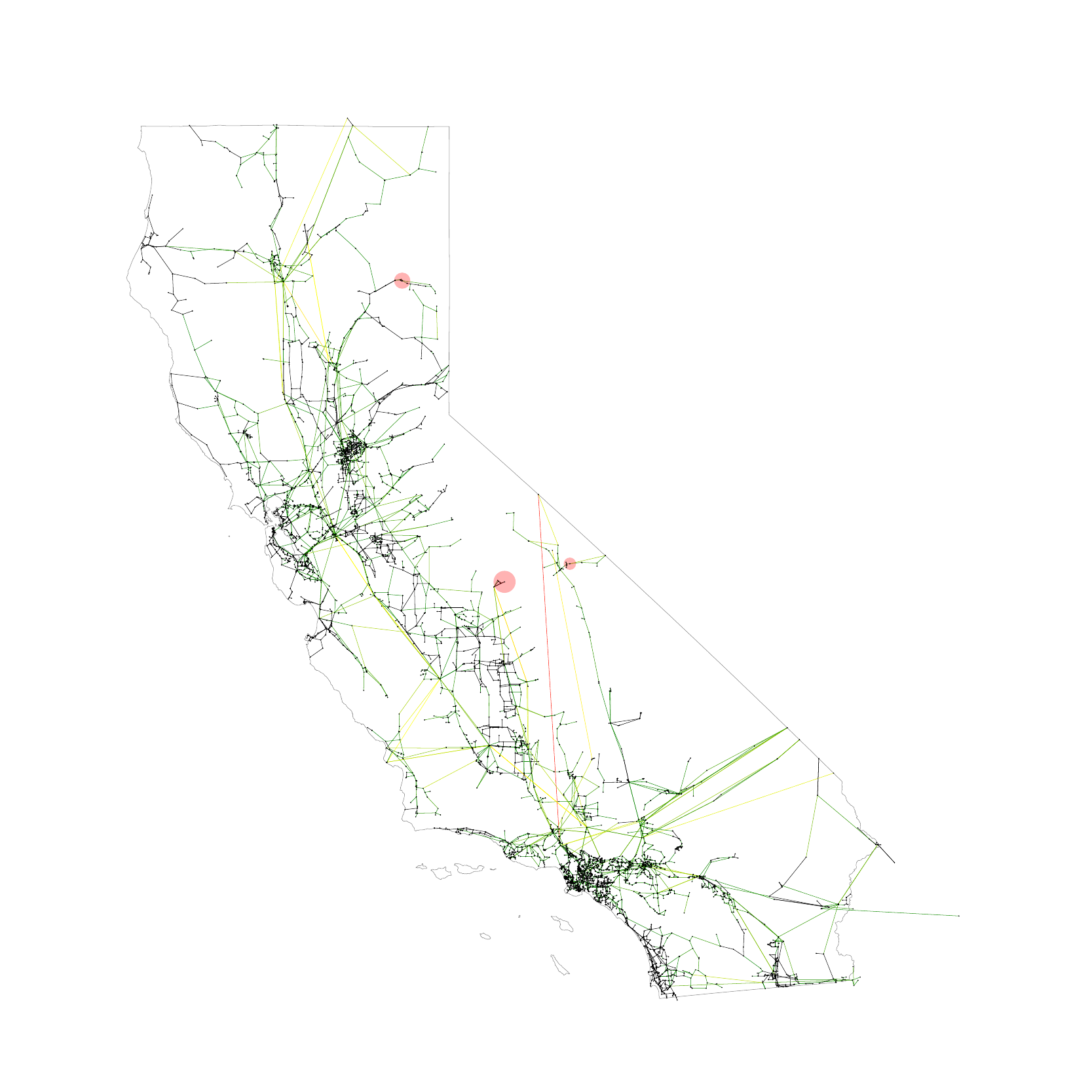}  
  \caption{Battery only placement results (Scheme~1) for the summer season. The size of the circle corresponds to the capacity of the installed battery with the largest circle equating to 400~MWh of installed capacity.}
  \label{subfig:summer_batts}
\end{subfigure}
\begin{subfigure}{.48\textwidth}
  \centering
  \captionsetup{width=.9\linewidth}
  \includegraphics[width=.99\linewidth]{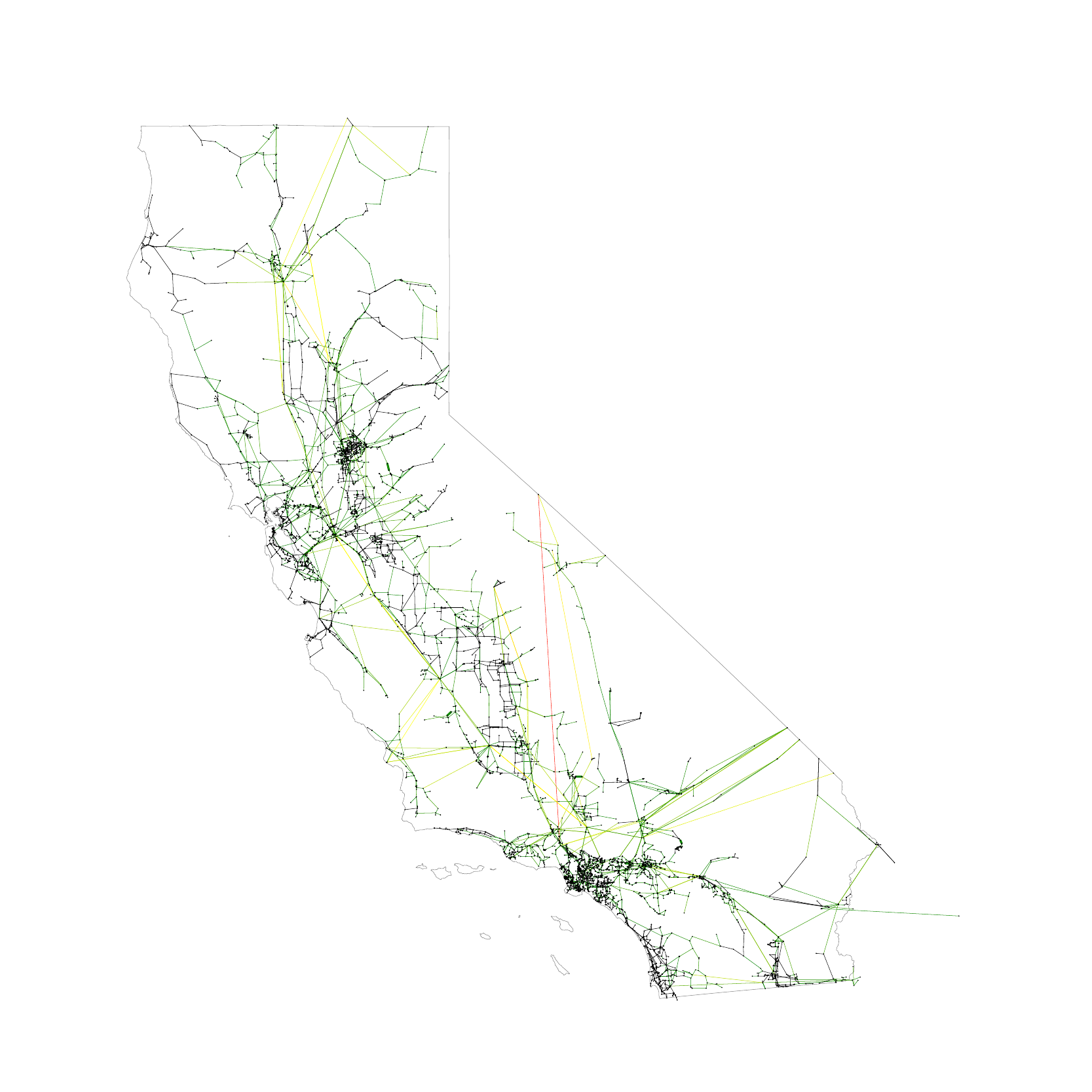}  
  \caption{Battery and undergrounding placement results (Scheme~2) for the summer season. Here, no batteries are installed. Thicker lines correspond to transmission lines that have been selected for undergrounding.}
  \label{subfig:summer_batts_ug}
\end{subfigure}
\caption{Optimal battery placements on the CATS network in Summer. Red circles are sized proportionally to the number of batteries placed at that bus with the largest circle representing 400~MWh of installed capacity. Thicker lines represent undergrounded transmission lines. Black lines represent transmission lines that always have zero risk values during the simulated scenarios within the season. All remaining lines have a color representing their maximum risk across the scenarios, with green lines representing lower risk and red lines representing higher risk.}
\label{figure:summer_results}
\end{figure*}

We run the planning problem across representative seasons, each containing three representative days, each from a different month. In addition, we run a planning problem across the full year, consisting of 12 representative days, one from each month. We consider the following seasonal breakdown:
\begin{itemize}
    \item \textbf{Spring:} March, April, May
    \item \textbf{Summer:} June, July, August
    \item \textbf{Fall:} September, October, November
    \item \textbf{Winter:} January, February, December
\end{itemize}
This results in each seasonal scenario having 3 subproblems while the full year scenario has 12 subproblems, one per representative month.

We next present the results considering two planning schemes:
\begin{itemize}
    \item \textbf{Scheme 1}: Batteries are the only options considered. All nodes in the network are considered potential candidates for battery placement.
    \item \textbf{Scheme 2}: Both batteries and line undergrounding are considered as investment options. All nodes in the network are considered as potential candidates for battery placement, and a set of risky lines is defined to indicate potential candidates for line undergrounding. We let any line $ij$ be in the set $\mathcal{L}^{\text{risk}}$ if it has a non-zero risk value for any of the scenarios within a considered season as described in Section~\ref{sec:case}. Specifically, the candidates for line undergrounding are obtained by the union of risky lines in all scenarios ($\mathcal{L}^\text{risk} = \bigcup_{\omega \in \Omega} \mathcal{L}_{\omega}\setminus\mathcal{L}_{\omega}^{\text{on}}$). According to this criterion, we obtain $|\mathcal{L}^\text{risk}| = 2103$, or 2103 line undergrounding candidates for the year-round scenario. Each candidate line corresponds to an existing connection in the network than can be upgraded (undergrounded) to remove the ignition risk while maintaining power flow.
\end{itemize}
Each problem is run until a 20 iteration limit.

\subsection{Seasonal results}
The optimized results (battery sizes and number of underground lines) are summarized in Table~\ref{tab:optimal_size}. This table presents the results for four different seasons. 
As shown in the table, we observe substantial battery placement in the case of Scheme~1 and note that the Spring season scenario sites the highest total battery size.

Comparing the results between the two planning schemes (Schemes~1 and~2) in Table~\ref{tab:optimal_size}, we observe that the planning scheme prefers line undergrounding over battery installation, as in Scheme~2, we observe that battery sizes are zero in most seasons compared to the results of Scheme~1. The authors found a similar preference for line undergrounding in~\cite{kody2022optimizing}.

\begin{table}[!htbp]
    \centering
    \caption{Optimized battery size (in p.u.) and number of lines undergrounded for different seasons (1 p.u. = 100 MWh).}
    \begin{tabular}{|c||c||c|c|}
    \hline 
    & \bf Scheme 1 & \multicolumn{2}{c|}{\bf Scheme 2 }\\ 
     & \bf (Battery only) & \multicolumn{2}{c|}{\bf (Battery + undergrounding) }\\ 
    \hline 
    & Battery [p.u.] & Battery [p.u.] & \# Lines undergrounded \\
    \hline\hline
    \bf  Spring  & 15.55 &  \hphantom{0}0.0\hphantom{0}  & 30 \\
     \hline
    \bf  Summer  & \hphantom{0}1.48 &  \hphantom{0}0.0\hphantom{0}  & \hphantom{0}5 \\
     \hline 
     \bf Fall   & \hphantom{0}6.82 &  26.14 & 25 \\
     \hline
    \bf  Winter  & 13.87 &  20.0\hphantom{0} & 59 \\
     \hline 
     \bf  Full Year  & 18.99 &  \hphantom{0}8.0\hphantom{0} & 99 \\
     \hline 
    \end{tabular}
    \label{tab:optimal_size}
\end{table}


\begin{figure*}[!htbp]
\centering
\begin{subfigure}{.48\textwidth}
  \centering
  \captionsetup{width=.9\linewidth}
  \includegraphics[width=.99\linewidth]{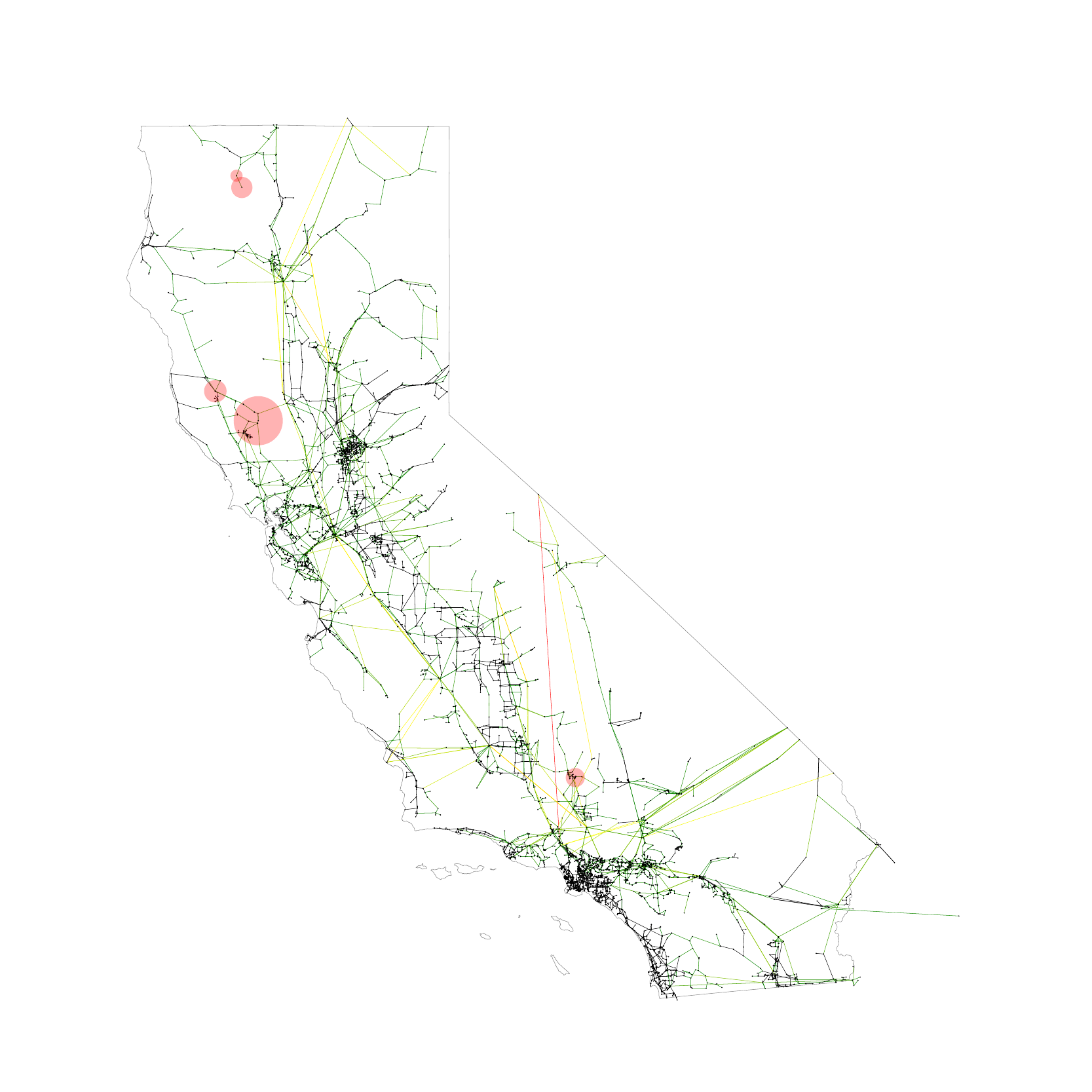}  
  \caption{Battery-only placement results (Scheme~1) for the fall season. The size of the circle corresponds to the capacity of the installed battery with the largest circle equating to 400~MWh of installed capacity.}
  \label{subfig:fall_batts}
\end{subfigure}
\begin{subfigure}{.48\textwidth}
  \centering
  \captionsetup{width=.9\linewidth}
  \includegraphics[width=.99\linewidth]{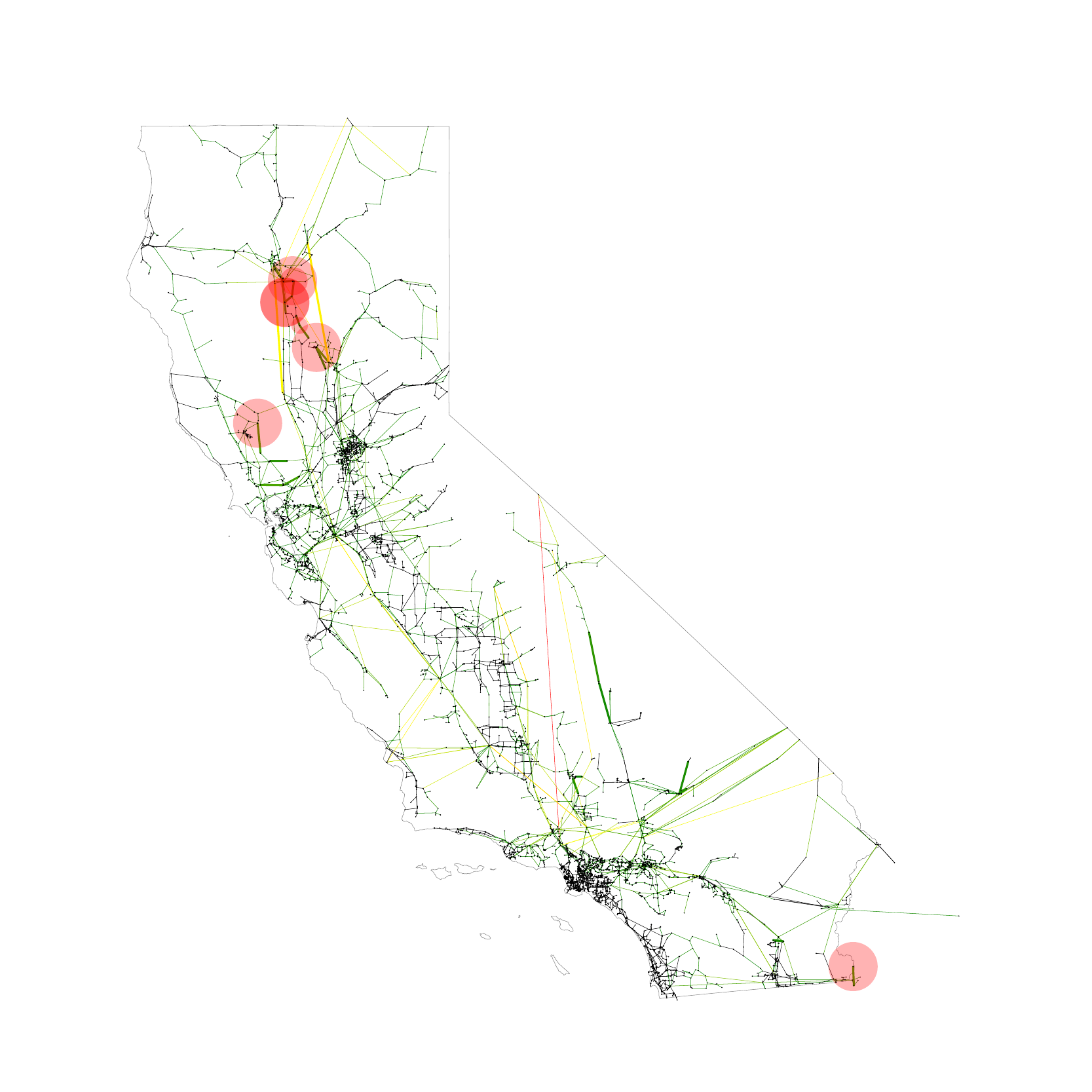}  
  \caption{Battery and undergrounding placement results (Scheme~2) for the fall season. Thicker lines correspond to transmission lines that have been selected for undergrounding.\newline}
  \label{subfig:fall_batts_ug}
\end{subfigure}
\caption{Optimal battery placements on the CATS network in Fall. Red circles are sized proportionally to the number of batteries placed at that bus with the largest circle representing 400~MWh of installed capacity. Thicker lines represent undergrounded transmission lines. Black lines represent transmission lines that always have zero risk values during the simulated scenarios within the season. All remaining lines have a color representing their maximum risk across the scenarios, with green lines representing lower risk and red lines representing higher risk.}
\label{figure:fall_results}
\end{figure*}

\begin{figure*}[!htbp]
\centering
\begin{subfigure}{.48\textwidth}
  \centering
  \captionsetup{width=.9\linewidth}
  \includegraphics[width=.99\linewidth]{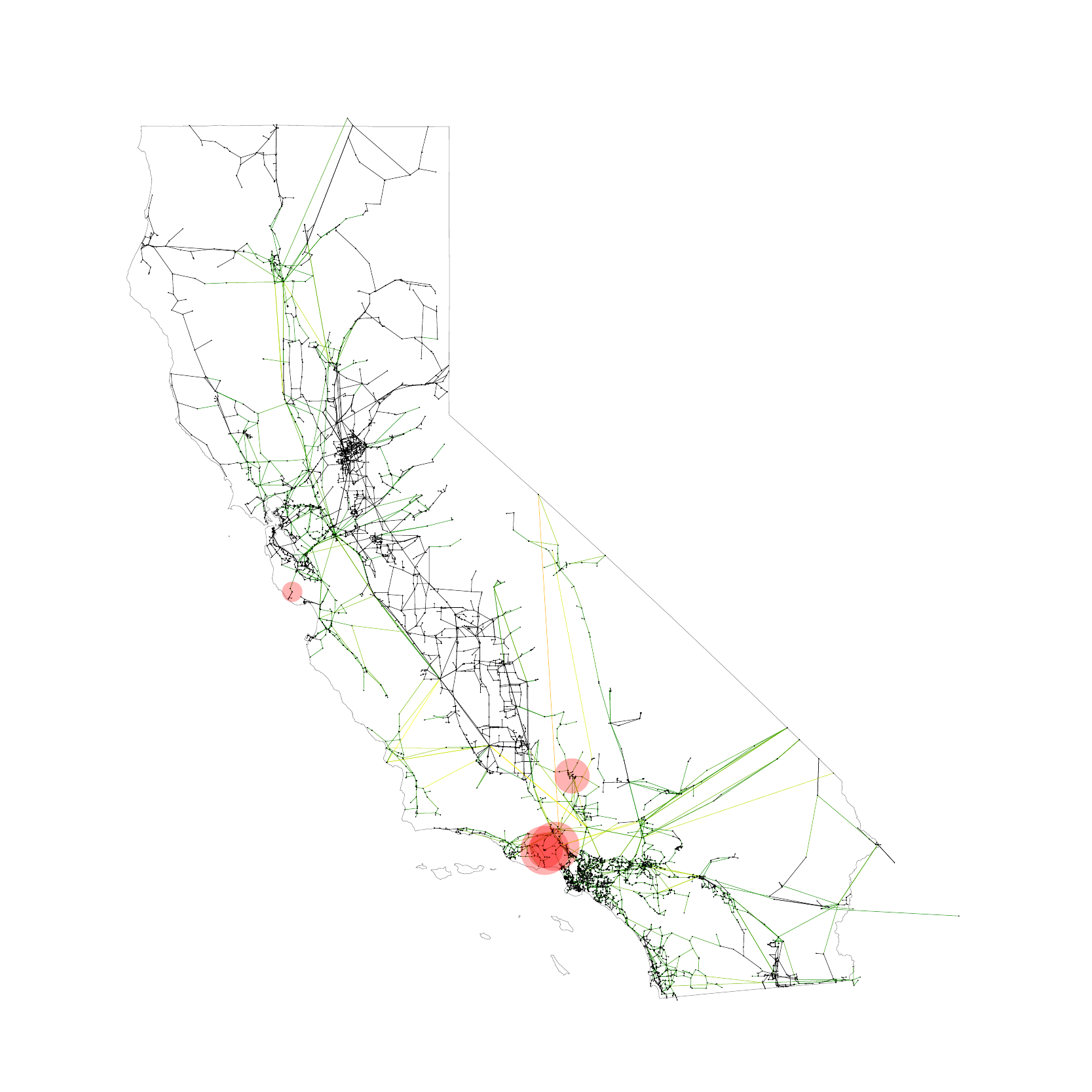}  
  \caption{Battery-only placement results (Scheme~1) for the winter season. The size of the circle corresponds to the capacity of the installed battery with the largest circle equating to 400~MWh of installed capacity.}
  \label{subfig:winter_batts}
\end{subfigure}
\begin{subfigure}{.48\textwidth}
  \centering
  \captionsetup{width=.9\linewidth}
  \includegraphics[width=.99\linewidth]{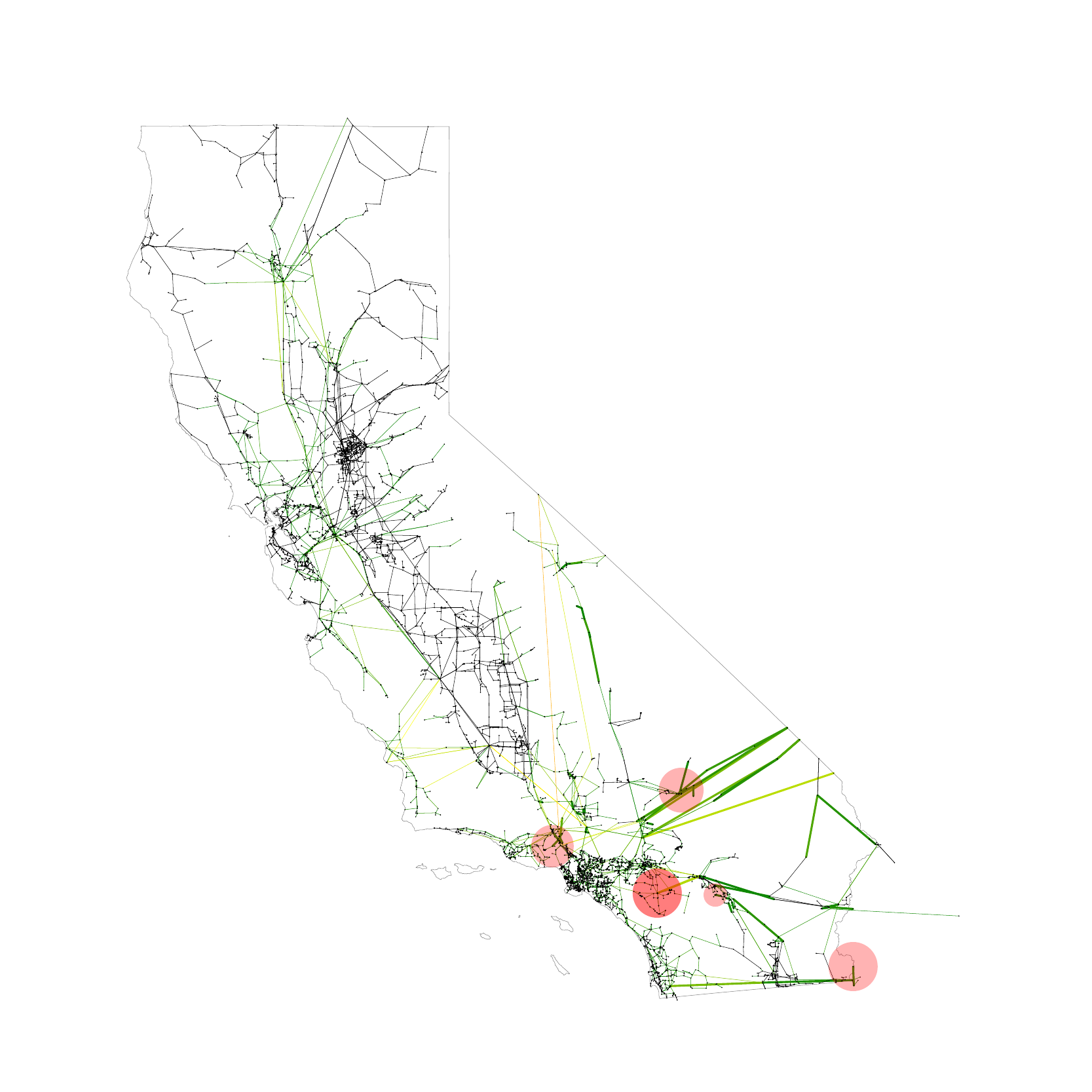}  
  \caption{Battery and undergrounding placement results (Scheme 2) for the winter season. Here, no batteries are installed. Thicker lines correspond to transmission lines that have been selected for undergrounding.}
  \label{subfig:winter_batts_ug}
\end{subfigure}
\caption{Optimal battery placements on the CATS network in Winter. Red circles are sized proportionally to the number of batteries placed at that bus with the largest circle representing 400~MWh of installed capacity. Thicker lines represent undergrounded transmission lines. Black lines represent transmission lines that always have zero risk values during the simulated scenarios within the season. All remaining lines have a color representing their maximum risk across the scenarios, with green lines representing lower risk and red lines representing higher risk.}
\label{figure:winter_results}
\end{figure*}

The corresponding results for load shedding with and without planning are summarized in Table~\ref{tab:load_shed}. As can be observed, the algorithm achieves a better reduction in the load shedding in Scheme 2, i.e., when we allow line undergrounding. This suggests that only considering battery placement planning may be insufficient for reducing the load shedding due to PSPS events. 

The corresponding placement decisions of the batteries and underground lines are shown in 
Figs.~\ref{figure:spring_results}, \ref{figure:summer_results}, \ref{figure:fall_results}, and \ref{figure:winter_results} for Schemes~1 (left) and~2 (right).
As can be seen in Fig.~\ref{subfig:spring_batts}, the spring season installs the most batteries when compared to the summer (Fig.~\ref{subfig:summer_batts}), fall (Fig.~\ref{subfig:fall_batts}), and winter (Fig.~\ref{subfig:winter_batts}) scenarios where only batteries can be placed (Scheme 1). When we look at the results where both batteries can be installed and the lines can be underground (Scehme 2), we observe that the fall and winter season install both simultaneously, as can be seen in Figs.~\ref{subfig:fall_batts_ug} and~\ref{subfig:winter_batts_ug} and Table~\ref{tab:optimal_size}. The summer season installs very few batteries (in the battery only case, Fig~\ref{subfig:summer_batts}) and undergrounds very few lines with no batteries (in the battery and undergrounding case, Fig~\ref{subfig:summer_batts_ug}). In Table~\ref{tab:load_shed}, we can see that the summer season has the largest amount of load shed. However, the Benders decomposition algorithm did not find a cost-effective investment outcome to reduce this load shed through battery installations or line undergrounding. In each of the four seasons, we do see very different investment decisions, motivating the need to plan more comprehensively with scenarios from all times of the year.

\begin{table*}[!htbp]
    \centering
    \caption{Per-scenario mean load shedding (in p.u.) without and with planning for different seasons (1 p.u. = 100 MWh).}
    \begin{tabular}{|c|c|c|c|c|c|c|c|}
    \hline 
    \multicolumn{2}{|c|}{} & \multicolumn{6}{c|}{\bf Load shedding mean per daily scenario}\\
    \hline
    Scenarios & Average Daily & \multicolumn{2}{c|}{Base case} & \multicolumn{2}{c|}{Scheme 1} & \multicolumn{2}{c|}{Scheme 2}\\ \cline{3-8}
     & Load (p.u.) & Absolute (p.u.) & Percentage & Absolute (p.u.) & Percentage & Absolute (p.u.) & Percentage\\ 
    \hline\hline 
    \bf  Spring  & 2572.1 & \phantom{0}33.92 & \hphantom{0}1.31 & \phantom{0}32.36  & \hphantom{0}1.25 &\phantom{0}27.74  & \hphantom{0}1.07\\
     \hline
    \bf  Summer & 2851.9 & 398.86  & 13.99 & 398.72 & 13.98 & 393.55 & 13.8\\
     \hline 
     \bf Fall  & 2727.4 & 320.30 & 11.74 & 318.18 & 11.66 &  304.15 & 11.15\\
    \hline
    \bf  Winter & 2555.4 & 185.50  & \hphantom{0}7.26 & 184.5 & \hphantom{0}7.12 & 171.20 & \hphantom{0}6.69\\
     \hline 
     \bf  Full Year & 2676.8 & 234.65  & \hphantom{0}8.76 & 232.40& \hphantom{0}8.68 & 211.75 & \hphantom{0}7.91\\
     \hline 
    \end{tabular}
    \label{tab:load_shed}
\end{table*}

\begin{figure*}[!htbp]
    \begin{subfigure}{.48\textwidth}
        \centering
        \captionsetup{width=.9\linewidth}
        \includegraphics[width=.95\linewidth]{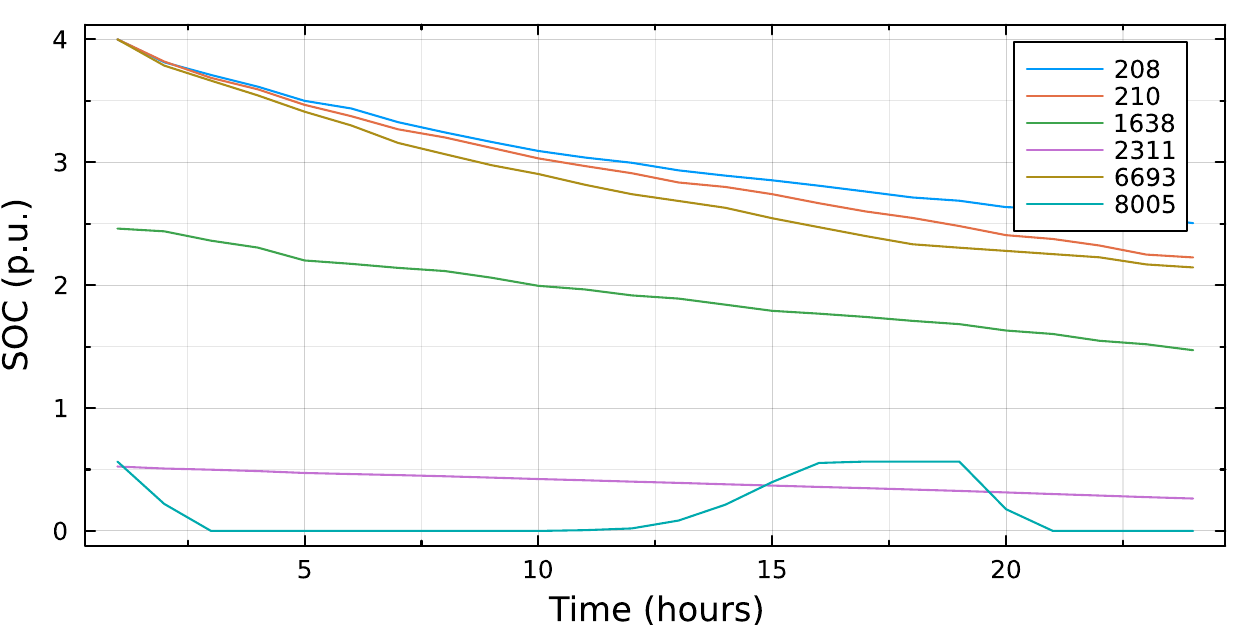} 
        \caption{SoE of batteries placed at the bus numbers indicated for the battery-only scheme in the spring season.}
        \label{subfig:spring_soe_batts}
    \end{subfigure}
    \begin{subfigure}{.48\textwidth}
        \centering
        \captionsetup{width=.9\linewidth}
        \includegraphics[width=.95\linewidth]{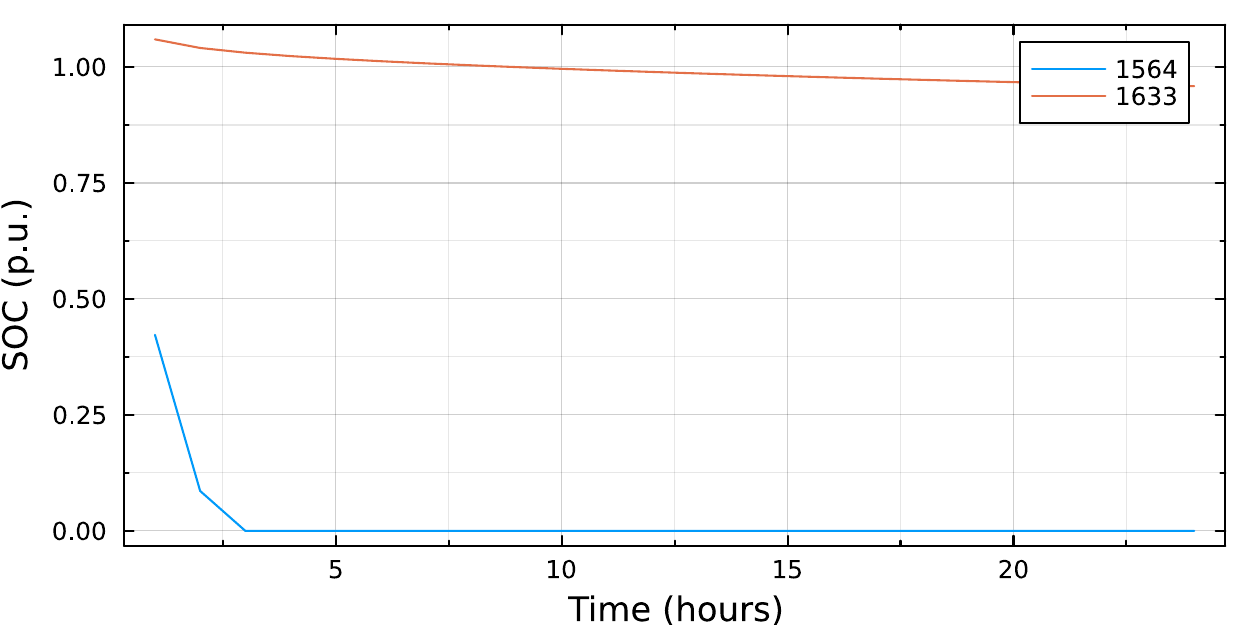}  
        \caption{SoE of batteries placed at the bus numbers indicated for the battery-only scheme in the summer season.}
        \label{subfig:summer_soe_batts}
    \end{subfigure}
    \\
    \begin{subfigure}{.48\textwidth}
        \centering
        \captionsetup{width=.9\linewidth}
        \includegraphics[width=.95\linewidth]{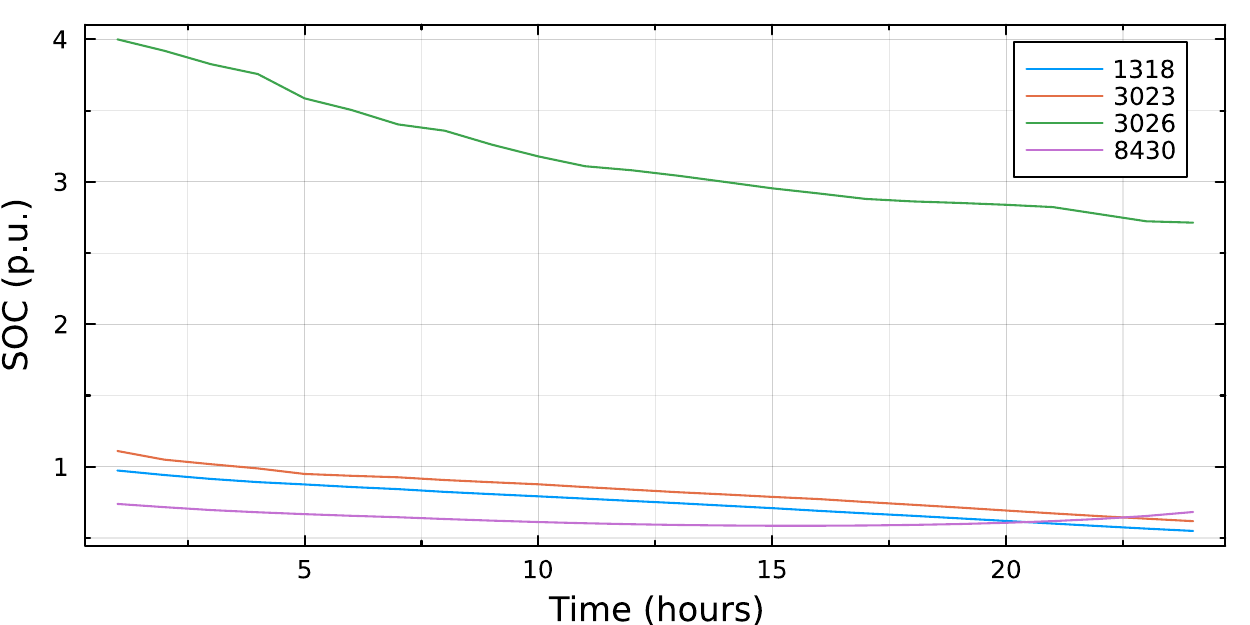}  
        \caption{SoE of batteries placed at the bus numbers indicated for the battery-only scheme in the fall season.}
        \label{subfig:fall_soe_batts}
    \end{subfigure}
    \begin{subfigure}{.48\textwidth}
        \centering
        \captionsetup{width=.9\linewidth}
        \includegraphics[width=.95\linewidth]{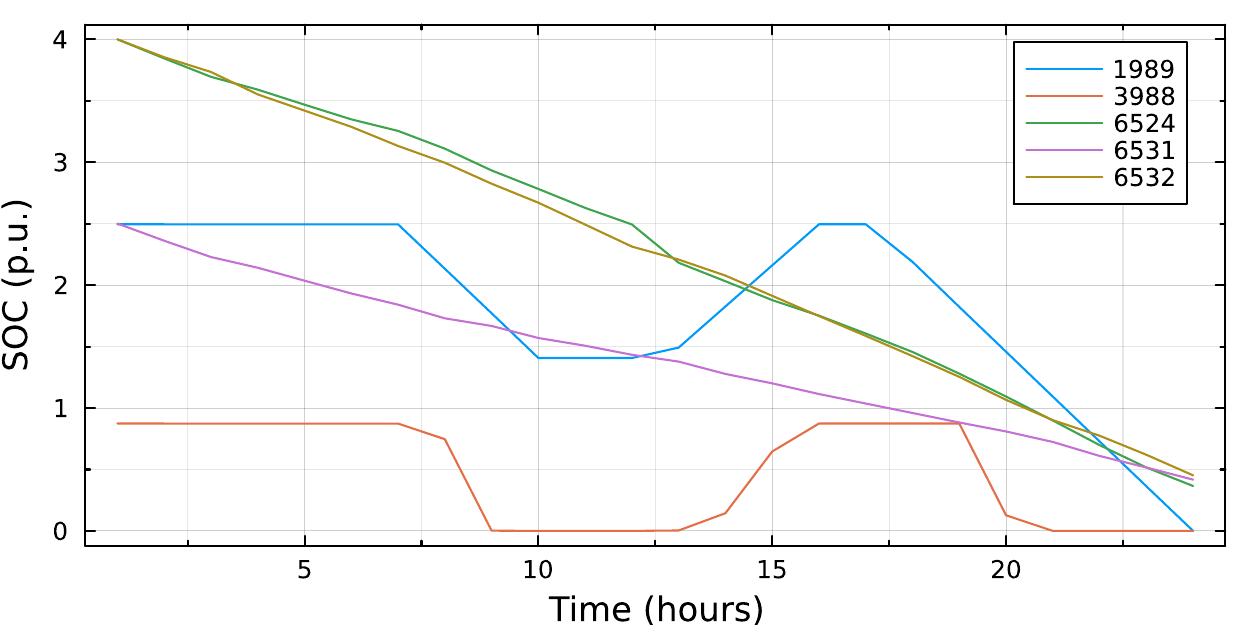} 
        \caption{SoE of batteries placed at the bus numbers indicated for the battery-only scheme in the winter season.}
        \label{subfig:winter_soe_batts}
    \end{subfigure}
    \caption{State of Energy (SoE) of the batteries placed within the battery-only scheme for the first 24 hour period of each season. Each line plot represented in the legend refer to the node index within the CATS network.}
     \label{fig:SOE_batts}
\end{figure*}

The State of Energy (SoE) for the sized batteries is shown in Fig.~\ref{fig:SOE_batts} for the four seasons. These results correspond to Scheme~1 (battery only). Given the lack of batteries placed in most seasonal scenarios in Scheme~2, we do not show SoE results from those installed batteries. As can be seen, the batteries mostly discharge during the day in the summer and fall seasons, seasons with relatively high wildfire ignition risk. During the spring and winter seasons, we observe that the batteries also charge during the day, which reflects a benefit for energy arbitrage purposes. Note that the days within each season are not linked together (i.e., the final time period of the first scenario does not have a state-of-energy constraint related to the starting time period of the second scenario). Accordingly, we show only the SoE results from the first scenario of each season. 

\subsection{Yearly results}
We also run a case for the full year, using the 12 days from each of the seasonal cases. The optimized result decisions (battery sizes and number of underground lines) are summarized in Table~\ref{tab:optimal_size}. As with the seasonal scenarios, we again see a greater reduction in load shed when considering both batteries and line undergrounding, as can be observed in Table~\ref{tab:load_shed}. While any investment serves to reduce load shed, the combination of line undergrounding and batteries achieves an additional reduction of 2065~MWh in daily average load shed in the full year compared to installing batteries alone.

\begin{figure*}[!htbp]
    \centering
    \includegraphics[width=0.99 \linewidth]{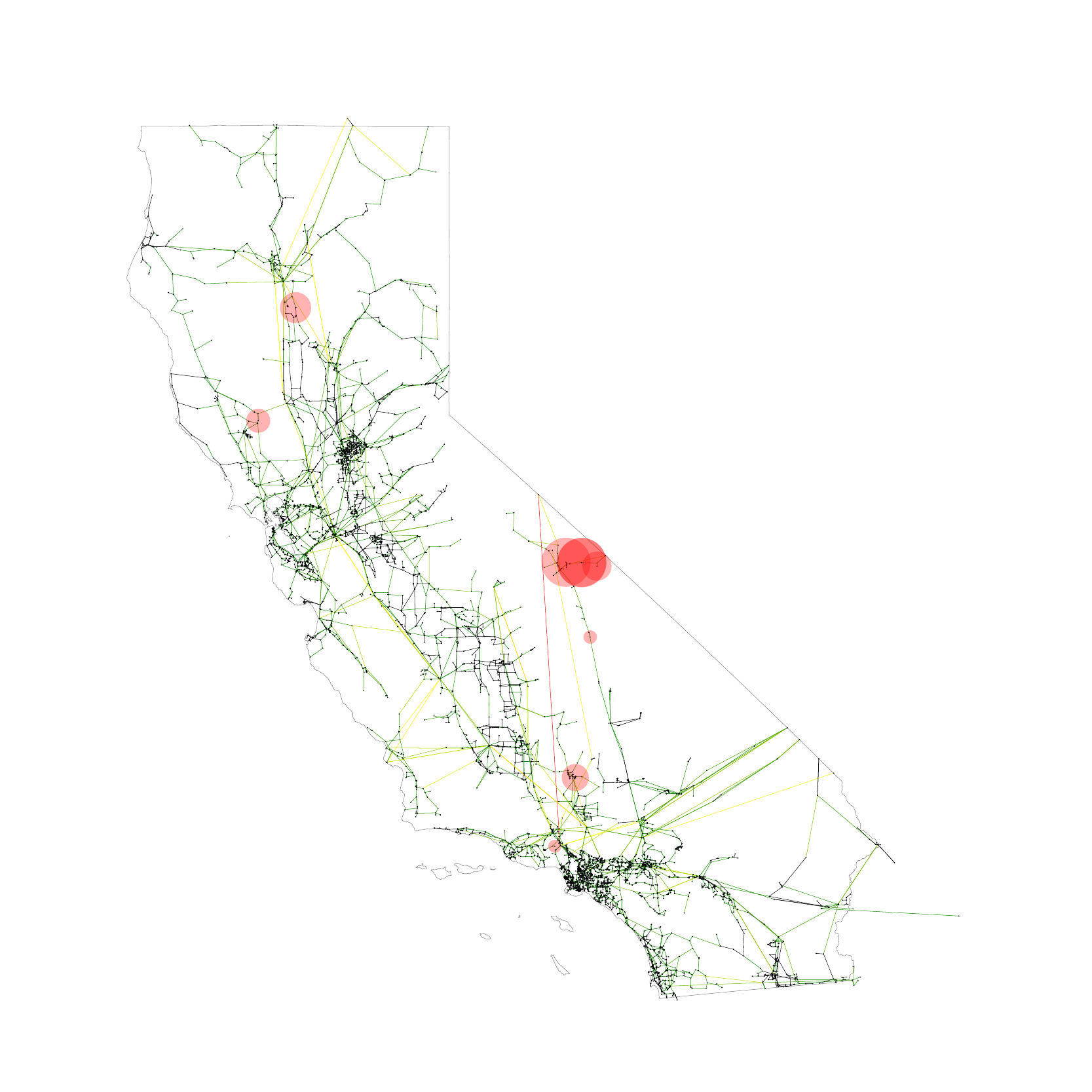}
    \caption{Optimal battery placements results (Scheme~1) on the CATS network for the full year, one scenario from each month. Red circles are sized proportionally to the number of batteries placed at that bus with the largest circle representing 400~MWh of installed capacity. Black lines represent transmission lines that always have zero risk values during the simulated scenarios within the year. All remaining lines have a color representing their maximum risk across the scenarios, with green lines representing lower risk and red lines representing higher risk.}
    \label{fig:full_results_batts}
\end{figure*}

\begin{figure*}[!htbp]
    \centering
    \includegraphics[width=0.99 \linewidth]{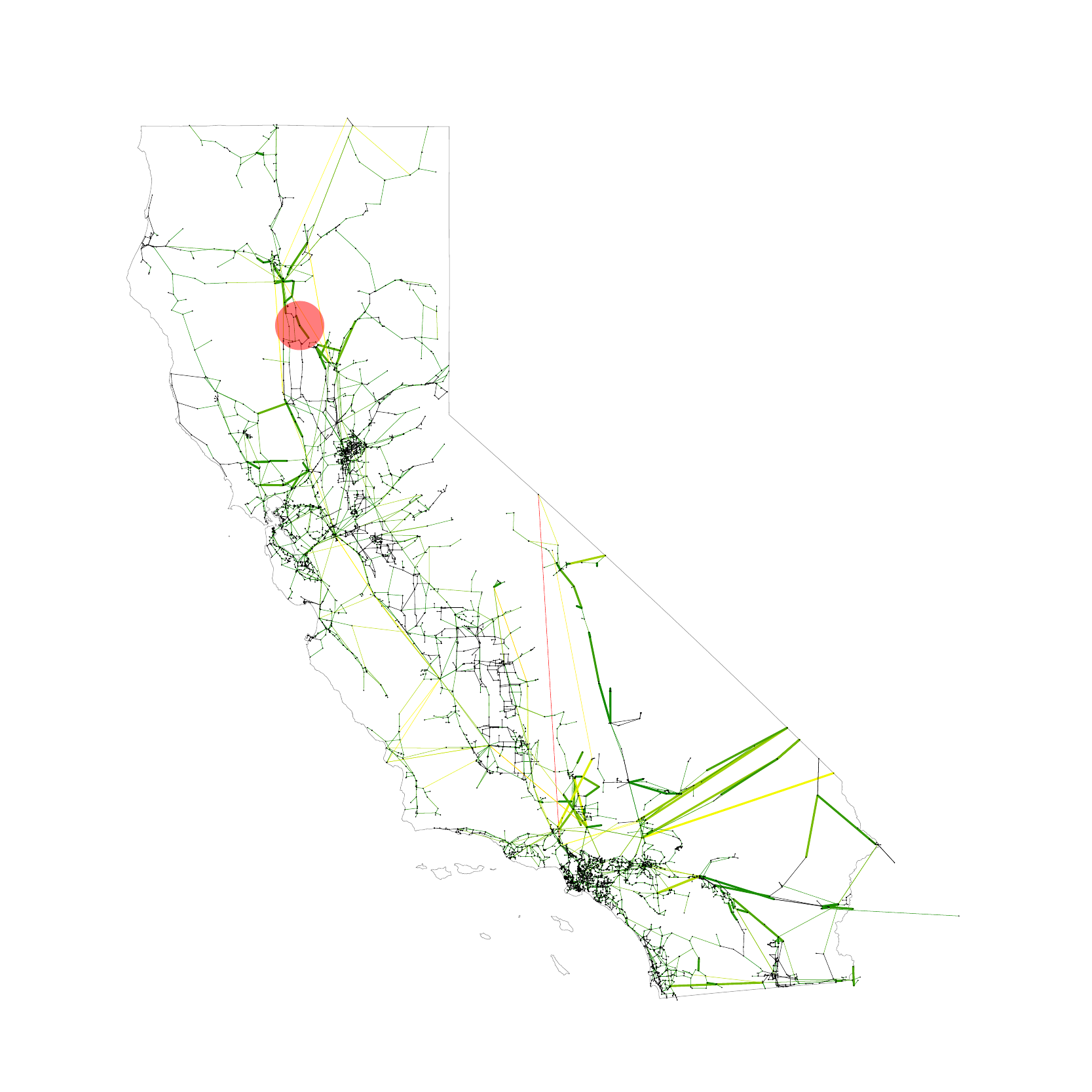}
    \caption{Optimal battery placements and line undergrounding decisions results (Scheme~2) on the CATS network for the full year, one scenario from each month. Red circles are sized proportionally to the number of batteries placed at that bus with the largest circle representing 400~MWh of installed capacity. Thicker lines represent undergrounded transmission lines. Black lines represent transmission lines that always have zero risk values during the simulated scenarios within the year. All remaining lines have a color representing their maximum risk across the scenarios, with green lines representing lower risk and red lines representing higher risk.}
    \label{fig:full_results_batts_ug}
\end{figure*}

The corresponding placement decisions of batteries and undergrounded lines are shown in Fig.~\ref{fig:full_results_batts} and Fig.~\ref{fig:full_results_batts_ug} for Schemes~1 and~2, respectively. Comparing the results between the seasonal and yearly simulations, we observe that the battery and line undergrounding placements are more geographically varied in the yearly results compared to the seasonal ones. Specifically, under investment Scheme~1, while there is some minor overlap between the full year results and spring/winter results, the full year results find optimal placements not selected in any of the seasonal results. Under Scheme~2, we see that the full year results install multiple batteries in addition to undergrounding transmission lines, something not seen in the seasonal results. This is expected as seasonal scenarios may not adequately capture the wildfire ignition risk across the transmission network's entire geographic region. This highlights the significance of incorporating many representative scenarios across the year into the planning process.

\begin{figure*}[!htbp]
    \begin{subfigure}{.98\textwidth}
        \centering
        \captionsetup{width=.9\linewidth}
        \includegraphics[width=.95\linewidth]{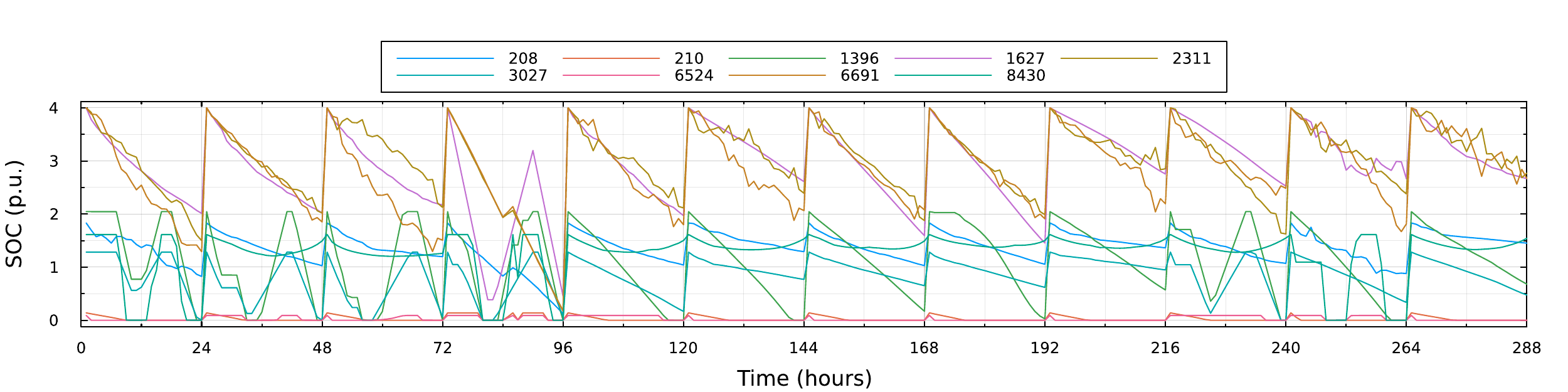} 
        \caption{SoE of batteries placed at the bus numbers indicated for the battery-only scheme for year-round scenarios.}
        \label{subfig:yearly_soe_batts}
    \end{subfigure}
    \begin{subfigure}{.98\textwidth}
        \centering
        \captionsetup{width=.9\linewidth}
        \includegraphics[width=.95\linewidth]{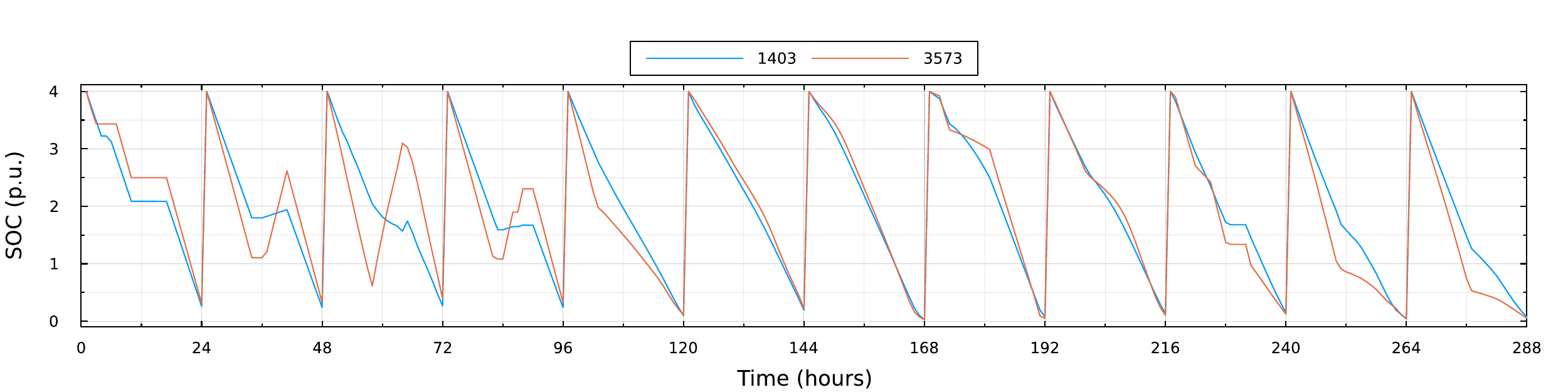}  
        \caption{SoE of batteries placed at the bus numbers indicated for the battery + undergrounding scheme for year-round scenarios.}
        \label{subfig:yearly_soe_batts_ug}
    \end{subfigure}
    \caption{State of Energy (SoE) of the batteries placed within the battery-only (Scheme~1) and battery + undergrounding (Scheme~2) schemes for the year round scenarios. The plots are shown for the 12 scenarios, where each 24-hour SoE profile is independent from the next 24-hour profile as the yearly scenario are modeled by distinct daily scenario representing each month (i.e., the final time period of scenario $\omega$ does not influence the first time period of $\omega+1$). Each line plot represented in the legend refer to the node index within the CATS network. }
     \label{fig:SOE_yearly}
     \vspace{-1em}
\end{figure*}

The State of Energy for the batteries sized for yearly scenarios are shown in Fig.~\ref{fig:SOE_yearly} for Schemes~1 and~2. The plots are shown for the 12 scenarios, where each 24-hour SoE profile is independent from the next 24-hour profile as the yearly scenarios are modeled by distinct daily scenarios representing each month (i.e., the final time period of scenario $\omega$ does not influence the first time period of $\omega+1$). Similar to the earlier SoE plots for the seasonal results, we observe that the batteries mostly discharge during the day in the summer and fall seasons (hours 120-263), seasons with relatively high wildfire ignition risk. During the spring and winter seasons (hours 0-119, 264-280), we observe that the batteries also charge during the day, which reflects a benefit for energy arbitrage purposes.

We also present the operational and investment costs for the yearly scenario for Schemes~1 and~2 in Table~\ref{tab:cost}. While the costs for batteries and undergrounding are large, these are upfront costs. The battery systems are modeled with a lifespan of 10 years~\cite{graf2022drives} and the undergrounded lines are modeled with a lifespan of 40 years~\cite{xcel2014overhead}. By dividing the upfront cost by the lifetime of the investment in days, we can look at a ``daily'' cost in Table~\ref{tab:cost_daily}. Here, we see an additional daily investment of 1.04 million USD in batteries saves 4.2 million USD in the cost of load shed from the results produced by Scheme~1. Likewise, a total average daily investment of 0.62 million in batteries and line undergrounding saves nearly 50 million in the daily average cost of load shed under Scheme~2.


\begin{table}[!htbp]
    \centering
    \caption{Total operational and investment cost for the 12-scenario yearly results (in millions US dollars).}
    \begin{tabular}{|c|c|c|c|}
    \hline 
    & \bf Baseline & \bf Scheme 1 & \bf Scheme 2 \\ 
     &  &(Battery only) & (Battery +  \\ 
          &  &  & undergrounding) \\ 
    \hline\hline 
    \textbf{Load shedding}  & 5630 & 5577 &  5082 \\
     \hline
    \textbf{Generation}  & 144 & \hphantom{0}141.8 &  \hphantom{0}139.3 \\
     \hline 
     \textbf{Battery}   & n.a. & 3800 &  1600 \\
     \hline
    \textbf{Undergrounding} & n.a. & n.a. &  2687 \\
     \hline 
    \end{tabular}
    \label{tab:cost}
\end{table}


\begin{table}[!htbp]
    \centering
    \caption{Operational and Investment cost from yearly scenario results, based on daily average (in millions US dollars).}
    \begin{tabular}{|c|c|c|c|}
    \hline 
    & \bf Baseline & \bf Scheme 1 & \bf Scheme 2 \\ 
     &  &(Battery only) & (Battery +  \\ 
          &  &  & undergrounding) \\ 
    \hline\hline 
    \textbf{Load shedding}  & 469 & 464.8 & 423.5 \\
     \hline
    \textbf{Generation}  & 12 & 11.8 &  11.6 \\
     \hline 
     \textbf{Battery}   & n.a. & 1.04 &  0.44 \\
     \hline
    \textbf{Undergrounding} & n.a. & n.a. &  0.18 \\
     \hline 
    \end{tabular}
    \label{tab:cost_daily}
\end{table}


\subsection{Computation}
We compare the solution time at the 20th iteration for a sequential implementation and a parallelized implementation in Table~\ref{tab:comp_time}.  All simulations are completed using Gurobi~11.0.1~\cite{gurobi} on the Partnership for an Advanced Computing Environment (PACE) at the Georgia Institute of Technology~\cite{PACE}. Seasonal scenarios are run on one node with 24 cores and 192 GB of memory. Each node has Dual Intel Xeon Gold 6226 CPUs @ 2.7 GHz. The full year results are run on nodes with larger amounts of memory. Each scheme was scheduled for different resources, namely 1.5 TB of memory for Scheme 1 and 768 GB of memory for Scheme 2. 
The sequential implementation solves each subproblem in order while the parallelized implementation solves each subproblem in parallel across cores in the node. The seasonal scenarios take about 13 hours to solve sequentially while the parallelized implementation solves in approximately 10 hours, yielding a nearly 25\% speed up. Alternatively, the full year scenarios takes 57 and 52 hours to solve sequentially for Schemes 1 and 2, respectively. The parallelized implementations solve in 25 and 19 hours for Schemes 1 and 2, respectively, for a roughly 60\% speed up. As mentioned, the full year scenarios were allocated different resources on a shared HPC cluster so the parallel Scheme 1 results were solved more slowly. The full year scenario, with four times as many subproblems, solve roughly twice as slowly  as the seasonal scenarios, indicating that this framework would facilitate extensions to more subproblems in future work.

\begin{table}[!htbp]
    \centering
    \caption{Per scenario solution time in minutes at the 20th iteration.}
    \begin{tabular}{|c|c|c|c|c|}
    \hline 
    & \multicolumn{4}{c|}{\bf Computation Time (in minutes)}\\
    \hline
    \multirow{2}{*}{Scenarios} & \multicolumn{2}{c|}{Scheme 1} & \multicolumn{2}{c|}{Scheme 2}\\ \cline{2-5}
     & Sequential & Parallelized & Sequential & Parallelized\\ 
    \hline\hline
    \bf  Spring  & \phantom{0}775  & 582 &\phantom{0}720  & 561\\
    \hline
    \bf  Summer & \phantom{0}796 & 566 & 745 & 550\\
    \hline 
    \bf  Fall  & \phantom{0}779 & 578 & 736 & 551\\
    \hline
    \bf  Winter & \phantom{0}784 & 582 & 732 & 582\\
    \hline 
    \bf  Full Year & 3418 & 1511 & 3130 & 1122\\
    \hline 
    
    \end{tabular}
    \label{tab:comp_time}
\end{table}

\section{Conclusion}\label{sec:conclusion}

In this paper, we implement a Benders decomposition algorithm to solve the battery sizing, siting, and operation problem, independently, and in tandem with line undergrounding investment decisions. This method allows for optimal investment decisions informed by the operation of the power grid from multiple scenarios with different associated real-world wildfire ignition risk, load demand, and renewable generation data. By incorporating scenarios from different times of the year, the investment decisions can aid in both load shed minimization as well as benefit the grid during nominal operations via actions like price arbitrage with the battery systems. Different scenarios provide greatly varying outcomes, as conditions shift throughout the year. By expanding the set of scenarios, we can install infrastructure to benefit operations throughout the year.

We found that when investments could contain both batteries and undergrounding, there was a preference to underground transmission lines. This could be due to the construction of the problem where large numbers of transmission lines are de-energized in PSPS events, incentivizing undergrounding lines so that they remain operational across multiple scenarios.


This method allows for an effective and scalable way to plan investments across multiple scenarios. Future work will investigate computational improvements for the parallelized operational subproblems. With these improvements, additional scenarios could be considered to enable infrastructure planning that more comprehensively accounts for the varying conditions throughout the year. Furthermore, including more climate impact data could facilitate more holistic modeling regarding the ability of infrastructure investments mitigate the impacts of many types of climate change driven natural disasters. Additionally, more sophisticated power flow models can be incorporated by leveraging both established and new AC power flow modeling techniques~\cite{fnt2019,haag2024,taheri_optimal_DCOTS}.




\section*{Acknowledgment}

The authors would like to thank Gurobi for the use of an academic license. The authors acknowledge support from the NSF AI Institute for Advances in Optimization (AI4OPT), \#2112533.


\bibliographystyle{IEEEtran}
\bibliography{IEEEabrv,refs.bib}

\end{document}